\documentclass[aps,prd,twocolumn,preprintnumbers, groupedaddress,nofootinbib,amssymb,notitlepage,eqsecnum]{revtex4-2}
\usepackage{bm}
\usepackage{amsmath,amsthm,amssymb}
\usepackage{amsfonts}
\usepackage{hyperref}
\usepackage{graphicx}
\usepackage{color}

\allowdisplaybreaks[1]

\newcommand{\be}{\begin{equation}}
\newcommand{\ee}{\end{equation}}
\newcommand{\ba}{\begin{eqnarray}}
\newcommand{\ea}{\end{eqnarray}}
\newcommand{\Mpl}{M_{\rm Pl}}

\begin{document}

\preprint{YITP-25-111, WUCG-25-08}

\title{Instability of regular 
planar black holes in four dimensions \\
arising from an infinite sum of curvature corrections}

\author{Antonio De Felice$^1$}

\author{Shinji Tsujikawa$^2$}

\affiliation{
$^1$Center for Gravitational Physics and Quantum Information, 
Yukawa Institute for Theoretical Physics, Kyoto University, 
606-8502, Kyoto, Japan}

\affiliation{
$^2$Department of Physics, Waseda University, 
3-4-1 Okubo, Shinjuku, Tokyo 169-8555, Japan}

\date{\today}

\begin{abstract}

In four-dimensional scalar-tensor theories 
derived via dimensional regularization with a 
conformal rescaling of the metric, we study the 
stability of planar black holes (BHs) whose horizons 
are described by two-dimensional compact Einstein 
spaces with vanishing curvature.  
By taking an infinite sum of Lovelock curvature invariants, 
it is possible to construct BH solutions whose metric components remain nonsingular at $r=0$, with a scalar-field derivative 
given by $\phi'(r)=1/r$, where $r$ is the radial 
coordinate. We show that such BH solutions suffer from a strong coupling problem, where the kinetic term of the even-parity scalar-field perturbation associated with the 
timelike coordinate vanishes everywhere.
Moreover, we find that these BHs are subject to both 
ghost and Laplacian instabilities for odd-parity 
perturbations near $r=0$. 
Consequently, the presence of these pathological features rules out regular planar BHs with the scalar-field profile 
$\phi'(r)=1/r$ as physically viable and stable configurations.

\end{abstract}

%\pacs{04.50.Kd,95.30.Sf,98.80.-k}

\maketitle

%%%%%%%%%%%%%%%%%%%%%%%%%%%%%
\section{Introduction}
\label{Intro}
%%%%%%%%%%%%%%%%%%%%%%%%%%%%%

The existence of black holes (BHs) is ubiquitous in 
general relativity (GR) and extended theories of gravity. 
In four spacetime dimensions, the vacuum solution of GR 
on a spherically symmetric and static (SSS) background 
is known as the Schwarzschild geometry \cite{Schwarzschild:1916uq}. 
The Schwarzschild solution has a curvature singularity at the center ($r=0$), where the Kretschmann scalar diverges. 
A similar singular behavior also occurs for charged 
BHs \cite{Reissner:1916cle} and rotating BHs \cite{Kerr:1963ud} in GR. 
This is attributed to Penrose’s singularity theorem in GR \cite{Penrose:1964wq}, which states that, under certain assumptions about spacetime and matter, singularities at the centers of BHs naturally arise as a consequence of 
gravitational collapse. 

In the presence of specific matter sources, numerous approaches have been proposed to construct nonsingular BH solutions in four dimensions. One well-known example is nonlinear electrodynamics (NED) within GR.
The NED Lagrangian ${\cal L}$ is a nonlinear 
function of the electromagnetic field strength  
$F=-F_{\mu \nu}F^{\mu \nu}/4$, where 
$F_{\mu \nu}=\partial_{\mu}A_{\nu}
-\partial_{\nu}A_{\mu}$ is the Maxwell tensor 
associated with a vector field $A_{\mu}$. 
With suitable choices of the functional form of 
${\cal L}(F)$, it is possible to construct 
SSS charged BH solutions that are 
free from curvature singularities at their centers \cite{Ayon-Beato:1998hmi,Ayon-Beato:2000mjt,Bronnikov:2000vy,Dymnikova:2004zc}. 
However, such nonsingular BHs are ruled out by 
angular Laplacian 
instabilities of vector-field perturbations near 
their centers \cite{DeFelice:2024seu,DeFelice:2024ops}.

If we consider a canonical scalar field $\phi$ 
minimally coupled to gravity on the SSS background, the resulting solution is the Schwarzschild BH without scalar hair \cite{Hawking:1971vc,Bekenstein:1972ny}.
This no-hair property holds for a broad class of 
scalar-tensor theories---including k-essence \cite{Graham:2014mda}, 
nonminimally coupled scalar fields with 
the Ricci scalar $R$ of the form 
$G_4(\phi)R$ \cite{Hawking:1972qk,Bekenstein:1995un,Sotiriou:2011dz}, 
and covariant Galileons \cite{Hui:2012qt,Babichev:2016rlq,Minamitsuji:2022mlv}. 
A notable exception is the Gauss-Bonnet (GB) curvature invariant
$R_{\rm GB}^2$ coupled to the scalar field 
in the form $\xi(\phi)R_{\rm GB}^2$ \cite{Kanti:1995vq,Torii:1996yi,Kanti:1997br,Sotiriou:2013qea,Doneva:2017bvd,Silva:2017uqg,Antoniou:2017acq,Minamitsuji:2018xde,Minamitsuji:2022mlv,Minamitsuji:2022vbi}.
In this case, the background geometry is modified from the Schwarzschild solution by the presence of a nonvanishing 
scalar-field profile, but a curvature singularity still remains at the center of BHs.

From a geometric perspective, the construction of
regular BHs in spacetime dimensions $d \geq 5$ 
has been investigated.
In the context of quasi-topological gravity \cite{Oliva:2010eb,Myers:2010ru,Dehghani:2011vu,Ahmed:2017jod,Cisterna:2017umf}, the possibility of eliminating the singularities of higher-dimensional BHs has been explored by incorporating an infinite tower of curvature corrections \cite{Bueno:2024dgm,Konoplya:2024hfg,Estrada:2024uuu,Frolov:2024hhe,Bueno:2024eig,Bueno:2024zsx}. 
In general, this geometric approach induces derivative terms higher than second order, raising concerns about the emergence of Ostrogradsky instabilities \cite{Ostrogradsky:1850fid}. 
Indeed, in some related theories like Einsteinian cubic 
gravity \cite{Bueno:2016xff} 
or Weyl gravity \cite{Stelle:1976gc}, it is known that 
the presence of higher-order derivative terms can give rise to 
ghost or Laplacian 
instabilities \cite{DeFelice:2023vmj,BeltranJimenez:2023mxp,DeFelice:2023kpl}.
Moreover, the above prescription of quasi-topological gravity 
is applicable only in spacetime dimensions $d \geq 5$ \cite{Bueno:2022res}.

In four-dimensional spacetime ($d=4$), there 
have been attempts to extract higher-curvature corrections by employing various 
regularization schemes \cite{Glavan:2019inb,Fernandes:2020nbq,Hennigar:2020lsl,Lu:2020iav,Colleaux:2020wfv,Fernandes:2025fnz} 
(see also Ref.~\cite{Kunstatter:2015vxa}).
If we restrict ourselves to local gravitational theories involving derivatives of the metric tensor $g_{\mu \nu}$ up to second order in order to avoid Ostrogradsky instabilities, the only nonvanishing Lagrangians in four dimensions are the cosmological constant $\Lambda$ and the Ricci scalar $R$  \cite{Lovelock:1971yv,Lovelock:1972vz}. 
The GB term $R_{\rm GB}^2$ is a specific 
combination of quadratic curvature invariants that leads to second-order field equations in arbitrary dimensions, but its contribution to the equations of motion identically 
vanishes in four dimensions. 
By rescaling the GB coupling constant $\alpha$ 
to $\alpha/(d-4)$ and then taking the limit $d \to 4$, 
one can extract the contribution 
of the GB term 
in four dimensions \cite{Glavan:2019inb}. 
However, this 4-dimensional 
Einstein-Gauss-Bonnet (4DEGB) gravity is 
plagued by several problems---such as the 
divergences in the perturbation 
equations \cite{Arrechea:2020evj,Arrechea:2020gjw}
and the absence of covariant equations 
for a graviton \cite{Gurses:2020ofy,Gurses:2020rxb}.

An alternative approach to regularization in 
4DEGB gravity is to perform a conformal rescaling of 
the metric tensor, $\tilde{g}_{\mu \nu}=e^{-2\phi} 
g_{\mu \nu}$, where $\phi$ is a scalar degree 
of freedom \cite{Fernandes:2020nbq,Hennigar:2020lsl}. 
The resulting four-dimensional action 
belongs to a subclass of shift-symmetric Horndeski 
theories \cite{Horndeski:1974wa,Deffayet:2011gz} in which 
the coupling functions $G_{2,3,4,5}$ 
depend only on the field kinetic term 
$X=-(1/2)\nabla^{\mu}\phi \nabla_{\mu}\phi$, 
where $\nabla_{\mu}$ denotes the covariant derivative. 
There is also another way of regularization based on 
a Kaluza-Klein reduction of the higher-dimensional 
EGB theory \cite{Lu:2020iav,Kobayashi:2020wqy,Ma:2020ufk}.
For a maximally symmetric and spatially flat internal space whose size is characterized by a scalar field, the effective four-dimensional action coincides with that derived via conformal rescaling. 
In this scalar-tensor formulation of 4DEGB gravity, there exists a hairy BH solution on the SSS background \cite{Glavan:2019inb, Lu:2020iav, Kobayashi:2011nu, Charmousis:2011bf}. 
However, it suffers from a strong coupling problem 
associated with a vanishing kinetic term of 
the scalar-field perturbation \cite{Tsujikawa:2022lww}. 
A similar strong coupling problem also arises 
for the dynamics of perturbations on the isotropic 
cosmological background \cite{Kobayashi:2020wqy}.

The conformal regularization performed in 4DEDB gravity 
can be extended to the $n$-th order Lovelock 
curvature invariants \cite{Lovelock:1971yv,Lovelock:1972vz} 
defined by 
\be
{\cal R}^{(n)} \equiv \frac{1}{2^n} 
\delta^{\mu_1 \nu_1 \cdots \mu_n \nu_n}_{\alpha_1 
\beta_1 \cdots \alpha_n \beta_n} 
\prod_{i=1}^{n} {R^{\alpha_i \beta_i}}_{\mu_i \nu_i}\,,
\label{calRn}
\ee
where 
\be
\delta^{\mu_1 \nu_1 \cdots \mu_n \nu_n}_{\alpha_1 
\beta_1 \cdots \alpha_n \beta_n} 
\equiv n! \delta^{\mu_1}_{[\alpha_1} 
\delta^{\nu_1}_{\beta_1} \cdots 
\delta^{\mu_n}_{\alpha_n} \delta^{\nu_n}_{\beta_n]}\,,
\ee
is the generalized Kronecker delta, and 
${R^{\alpha_i \beta_i}}_{\mu_i \nu_i}$ is the Riemann 
tensor. For each integer value of $n \geq 0$, 
the Lagrangian ${\cal R}^{(n)}$ leads to 
second-order field equations of motion. 
We note that the integers $n=0,1,2$ 
correspond, respectively, to 
${\cal R}^{(0)}=1$, 
${\cal R}^{(1)}=R$, and 
${\cal R}^{(2)}=R_{\rm GB}^2$.
For the spacetime dimension $d>2n$, 
the field equations of motion acquire 
contributions from the $n$-th 
order Lagrangian ${\cal R}^{(n)}$.
As $d$ approaches the critical dimension $2n$, 
one can perform a conformal regularization by taking 
the following limit at each 
order $n$ \cite{Fernandes:2020nbq,Hennigar:2020lsl,Colleaux:2020wfv,Fernandes:2025fnz}: 
\be
{\cal L}^{(n)}=\lim_{d \to 2n} 
\frac{\sqrt{-g}{\cal R}^{(n)}-\sqrt{-\tilde{g}} 
\tilde{\cal R}^{(n)}}{d-2n}\,, 
\label{Ln0}
\ee
where a tilde denotes quantities defined in the 
conformally related frame with metric tensor 
$\tilde{g}_{\mu \nu}=e^{-2\phi} g_{\mu \nu}$. 

The conformal regularization in 4DEGB gravity 
corresponds to the case $n=2$ in Eq.~(\ref{Ln0}), 
yielding the Lagrangian ${\cal L}^{(2)}$ 
in the limit $d \to 4$.
Although we are working in four dimensions, it is still possible 
to include an infinite tower of Lagrangians ${\cal L}^{(n)}$ 
with $n=2,3,\ldots$, by taking the limit $d \to 2n$ 
at each order $n$ \cite{Colleaux:2020wfv,Fernandes:2025fnz}. 
In particular, summing over the infinite tower of ${\cal L}^{(n)}$ 
for all $n=2,3,\ldots$ leads to a subclass of shift-symmetric 
Horndeski theories in four dimensions.

If we apply such theories to the dynamics on an 
isotropic and homogenous cosmological background, the initial 
big-bang singularity can be replaced by an inflationary solution 
characterized by a finite Hubble expansion 
rate $H$ \cite{Fernandes:2025fnz}. 
The background solution in which the time derivative of 
the scalar field, $\dot{\phi}$, 
equals $H$ gives rise to this nonsingular inflationary cosmology. 
However, it was recently found that this solution 
suffers from a strong coupling problem for the scalar-field perturbation, as well as 
Laplacian instabilities in the tensor sector~\cite{Tsujikawa:2025eac}.

In four-dimensional theories containing an infinite sum of curvature corrections, there exist planar BH solutions that are 
regular at $r=0$ \cite{Fernandes:2025fnz}. The horizon of such BHs has a two-dimensional flat topology. 
In GR, planar BHs can be realized in an asymptotically 
Anti-de Sitter (AdS) spacetime with a negative 
cosmological constant \cite{Lemos:1994fn}. 
The BHs with a two-dimensional planar topology, which may be interpreted as black membranes, are related to cylindrical BHs (black strings) \cite{Lemos:1994xp} and to toroidal BHs (black tori) \cite{Lemos:1995cm} through the compactification of one and two coordinates, respectively. 
Planar BHs can form as a result of the gravitational 
collapse of a planar distribution of matter~\cite{Lemos:1997bd}. 
Ref.~\cite{Cardoso:2001vs} computed the quasinormal modes 
of planar, cylindrical, and toroidal BHs in GR, all of which 
share the same quasinormal frequencies.

In theories arising from an infinite sum of curvature corrections, 
the geometry of planar BHs is modified by the presence of a nonvanishing scalar-field radial derivative of the form $\phi'(r) = 1/r$~\cite{Fernandes:2025fnz}. 
We emphasize that this type of solution arises in planar 
two-dimensional geometry, whereas for standard spherically symmetric BHs with a two-dimensional closed geometry, 
the scalar-field solution takes a different form. 
In the latter case, no linearly stable, regular, spherically symmetric BHs with nontrivial scalar hair have been reported 
in the literature \cite{Minamitsuji:2022mlv,Minamitsuji:2022vbi}.

The metric components of planar BHs found in Ref.~\cite{Fernandes:2025fnz} 
remain regular at $r=0$, but it is not yet clear whether they are stable against 
linear perturbations. In this paper, we address the linear stability of planar BHs that are regular at $r=0$. Since the two-dimensional topology of these BHs 
differs from that of the SSS spacetime, 
it is necessary to formulate BH perturbations 
on a different background 
within the framework of shift-symmetric Horndeski theories.
We compute the second-order actions for both the odd- and even-parity 
sectors and derive the linear stability conditions 
for the three dynamical degrees of freedom: 
one scalar mode and two gravitational modes.
In the odd-parity sector, we show that the BH solutions 
with $\phi'(r)=1/r$ suffer from 
ghost and Laplacian instabilities near $r=0$.
Moreover, the kinetic term 
of the even-parity scalar-field perturbation vanishes 
both inside and outside the horizon.
As a result, BHs that are regular at $r=0$ are plagued 
by an infinitely strong coupling problem. 
Thus, these regular planar BHs 
are ruled out as stable configurations.

This paper is organized as follows. 
In Sec.~\ref{4Dsec}, we briefly review the four-dimensional 
action of regularized Lovelock gravity obtained 
through a conformal rescaling of the metric.
In Sec.~\ref{BHsec}, we investigate the properties of 
planar BHs that are free from singularities at $r=0$. 
In Sec.~\ref{persec}, we analyze perturbations of 
planar BHs and discuss the issue of gauge choices.
In Sec.~\ref{oddsec}, we derive the stability conditions of regular planar BHs under odd-parity perturbations and show that these solutions are generally subject to ghost and 
Laplacian instabilities.
In Sec.~\ref{evensec}, we turn to the analysis of even-parity perturbations 
and demonstrate that regular BHs with $\phi'(r)=1/r$ inevitably suffer from a strong coupling problem due to the vanishing kinetic term of the scalar-field perturbation.
Finally, Sec.~\ref{consec} is devoted to conclusions.

Throughout this paper, we use units in which the speed of light $c$, the reduced Planck constant $\hbar$, and the reduced Planck mass $\Mpl$ are all set to 1.

%%%%%%%%%%%%%%%%%%%%%%%%%%%%%%%%%%%%%%%%%%%%%%%%%%%%%%%
\section{Regularized four-dimensional Lovelock gravity}
\label{4Dsec}
%%%%%%%%%%%%%%%%%%%%%%%%%%%%%%%%%%%%%%%%%%%%%%%%%%%%%%%

As mentioned in Introduction, the Lagrangian
${\cal L}^{(n)}$ in Eq.~(\ref{Ln0}) arises from
dimensional regularization combined with a
conformal rescaling of the metric,
$\tilde{g}_{\mu \nu}=e^{-2\phi} 
g_{\mu \nu}$ \cite{Fernandes:2020nbq,Hennigar:2020lsl,Colleaux:2020wfv,Fernandes:2025fnz}.
There exists an infinite tower of curvature corrections
labeled by $n=2,3,\cdots$, where $n=2$ corresponds
to the case of 4DEGB gravity.
For $n \geq 3$, the spacetime dimension $d$
is in the range $d \geq 6$ when taking the limit $d \to 2n$.
Nevertheless, it is still possible to include 
the Lagrangians ${\cal L}^{(n)}$ with $n \geq 3$ 
in four spacetime dimensions
\cite{Colleaux:2020wfv,Fernandes:2025fnz}.
By summing over all ${\cal L}^{(n)}$ for $n=2,3,\cdots$, together with
the Ricci scalar $R$ and the cosmological constant $\Lambda$,
the resulting action is given by 
\be
{\cal S}=
\int {\rm d}^4x\sqrt{-g}  
\left[
\frac{1}{2} R -\Lambda + \frac{1}{2\ell^2} 
\sum_{n=2}^{\infty} c_n \ell^{2n} 
{\cal L}^{(n)}\right]\,,
\label{action}
\ee
where $g$ is the determinant of the metric tensor
$g_{\mu \nu}$, $\ell$ is a constant length scale,
and $c_n$ are constant coefficients. 
The Lagrangian ${\cal L}^{(n)}$ at each order 
$n$ takes the form 
\ba
{\cal L}^{(n)} &=& 
G_2^{(n)}(X)-G_{3}^{(n)}(X)\square\phi 
+G_{4}^{(n)}(X) R \nonumber \\
& &+G_{4,X}^{(n)}(X) \left[ (\square \phi)^{2}
-(\nabla_{\mu}\nabla_{\nu} \phi)
(\nabla^{\mu}\nabla^{\nu} \phi) \right] \nonumber \\
& &
+G_{5}^{(n)}(X) G_{\mu \nu} \nabla^{\mu}\nabla^{\nu} \phi
\nonumber \\
& &
-\frac{1}{6}G_{5,X}^{(n)}(X)
[ (\square \phi )^{3}-3(\square \phi)\,
(\nabla_{\mu}\nabla_{\nu} \phi)
(\nabla^{\mu}\nabla^{\nu} \phi) \nonumber \\
& &
+2(\nabla^{\mu}\nabla_{\alpha} \phi)
(\nabla^{\alpha}\nabla_{\beta} \phi)
(\nabla^{\beta}\nabla_{\mu} \phi) ]  \,,
\label{Ln}
\ea
where $G_{\mu \nu}$ is the Einstein tensor, 
and $G_{2,3,4,5}^{(n)}(X)$ are functions 
of $X=-(1/2)\nabla^{\mu}\phi \nabla_{\mu}\phi$, 
whose explicit forms are given by 
\ba
G_2^{(n)}(X) &=& 2^{n+1} (n-1)(2n-3)X^n\,,
\label{Gchoice0} \nonumber \\
G_3^{(n)}(X) &=& -2^n n (2n-3) X^{n-1}\,,\nonumber \\
G_4^{(n)}(X) &=& 2^{n-1} n X^{n-1}\,,\nonumber \\
G_5^{(n)}(X) &=& 
\begin{cases}
-4 \ln |X|& \quad ({\rm for}~n=2)\,, \nonumber \\
-2^{n-1} \dfrac{n(n-1)}{n-2} 
X^{n-2} & \quad ({\rm for}~n \geq 3)\,.
\end{cases}
\label{Gchoice}\\
\ea
In the action (\ref{action}), we have used 
the notations  
$\square=\nabla^{\mu} \nabla_{\mu}$ and 
$G_{i,X}^{(n)}(X)
=\partial G_i^{(n)}/\partial X$. 
The difference from the application to cosmological 
dynamics performed in Refs.~\cite{Fernandes:2025fnz,Tsujikawa:2025eac} 
is that matter fluids (such as radiation) are not 
included in Eq.~(\ref{action}).
By taking the infinite sum of $G_i^{(n)}$ over 
$n=2,3,\cdots$, the action~(\ref{action}) can be 
expressed in the form
\ba
{\cal S} &=& \int {\rm d}^4 x \sqrt{-g}\,
\{ G_2 (X)-G_{3} (X)\square\phi 
+G_{4} (X) R \nonumber \\
& &+G_{4,X}(X) \left[ (\square \phi)^{2}
-(\nabla_{\mu}\nabla_{\nu} \phi)
(\nabla^{\mu}\nabla^{\nu} \phi) \right] \nonumber \\
& &
+G_{5}(X) G_{\mu \nu} \nabla^{\mu}\nabla^{\nu} \phi
\nonumber \\
& &
-\frac{1}{6}G_{5,X} (X)
[ (\square \phi )^{3}-3(\square \phi)\,
(\nabla_{\mu}\nabla_{\nu} \phi)
(\nabla^{\mu}\nabla^{\nu} \phi) \nonumber \\
& &
+2(\nabla^{\mu}\nabla_{\alpha} \phi)
(\nabla^{\alpha}\nabla_{\beta} \phi)
(\nabla^{\beta}\nabla_{\mu} \phi) ] \}  \,,
\label{Ln2}
\ea
where
\ba
G_{2}(X) &= & 
-\Lambda+\frac{1}{2 \ell^2} 
\sum_{n=2}^{\infty} 
c_n \ell^{2n} G_2^{(n)}(X)\,,\label{G2} \\
G_{3}(X) &=& 
\frac{1}{2\ell^2}\sum_{n=2}^{\infty} 
c_n \ell^{2n} G_3^{(n)}(X)\,,\\
G_{4}(X) &=& \frac{1}{2} 
+ \frac{1}{2\ell^2}\sum_{n=2}^{\infty} 
c_n \ell^{2n} G_4^{(n)}(X) \,,
\label{G4} \\
G_{5}(X) &=&
\frac{1}{2\ell^2}\sum_{n=2}^{\infty} 
c_n \ell^{2n} G_5^{(n)}(X)\,.
\label{G5}
\ea
Thus, the effective four-dimensional action belongs to 
the framework of shift-symmetric Horndeski theories 
whose action is invariant under a constant shift of 
the scalar field, $\phi \to \phi+c$.

Depending on the choice of coefficients $c_n$, 
the resulting coupling functions $G_{2,3,4,5}(X)$ differ. 
For the choice
\be
c_n=1 \quad {\rm for~all~}n 
\qquad ({\rm Model~1})\,,
\ee
we obtain \cite{Fernandes:2025fnz}
\ba
\hspace{-1cm}
G_2(X) &=& -\Lambda +\frac{4 \ell^2 X^2 
(1+6 \ell^2 X)}{(1-2\ell^2 X)^3}, \nonumber \\
\hspace{-1cm}
G_3(X) &=& \frac{1-10 \ell^2 X}{(1-2\ell^2 X)^3}\,, \nonumber \\
\hspace{-1cm}
G_4(X) &=& \frac{1}{2(1-2\ell^2 X)^2}\,,\nonumber \\
\hspace{-1cm}
G_5(X) &=& -2 \ell^2 \left[ \frac{1-2\ell^4 X^2}
{(1-2\ell^2 X)^2}+\ln \left( \frac{2\ell^2 |X|}{1-2\ell^2 X} 
\right) \right].
\label{example1}
\ea
It should be noted that this action is well-defined only when 
$1-2\ell^2X>0$. If this condition is not satisfied, the last 
term of $G_5(X)$ can be generalized to the form 
$\ln[2\ell^2|X|/|1-2\ell^2 X|]$. In particular, $X$ is not generally bounded in this theory.
A discussion of this generalization of the Horndeski coupling functions is presented in Appendix~A.

If we choose
\be
c_n=\frac{1-(-1)^n}{2n} \qquad ({\rm Model~2})\,,
\ee
we have
\ba
G_2(X) &=& -\Lambda+\frac{2 X(28 \ell^4 X^2-3)}
{(1-4\ell^4 X^2)^2}+\frac{3}{\ell^2} 
\tanh^{-1} (2\ell^2 X), \nonumber \\
G_3(X) &=& -\frac{4 \ell^4 X^2 (3+4\ell^4 X^2)}
{(1-4\ell^4 X^2)^2}\,, \nonumber \\
G_4(X) &=& \frac{1}{2(1-4\ell^4 X^2)}\,,\nonumber \\
G_5(X) &=& -\frac{2 \ell^4 X}{1-4\ell^4 X^2}
-\ell^2 \tanh^{-1} (2\ell^2 X)\,.
\label{example2}
\ea
For other choices of $c_n$, the readers may refer to Refs.~\cite{Fernandes:2025fnz,Tsujikawa:2025eac}. The point is that it is possible to realize planar BH solutions without singularities at $r=0$ by taking an infinite sum over $n$. This issue will be addressed in Sec.~\ref{BHsec}.

%%%%%%%%%%%%%%%%%%%%%%%%%%%%%
\section{Planar BHs} 
\label{BHsec}
%%%%%%%%%%%%%%%%%%%%%%%%%%%%%

We consider the background line element 
given by
\be
{\rm d}s^2=-f(r) {\rm d}t^2
+h^{-1}(r){\rm d}r^2
+r^2 {\rm d} l^2\,,
\label{line}
\ee
where $f$ and $h$ are functions of the radial 
coordinate $r$, and ${\rm d} l^2={\rm d} x_2^2+{\rm d} x_3^2$. 
Unlike the SSS BH, where the two-dimensional 
line element ${\rm d} l^2$ is expressed as
${\rm d} l^2={\rm d}\theta^2
+\sin^2 \theta\, {\rm d}\varphi^2$, 
the line element (\ref{line}) has 
a two-dimensional flat topology 
with the coordinate $(x_2,x_3)$.
We consider a radial-dependent background 
scalar field, $\phi=\phi(r)$.
The scalar kinetic term on the 
background (\ref{line}) 
is given by 
\be
X=-\frac{1}{2}h \phi'^2\,,
\ee
where a prime denotes the derivative 
with respect to $r$.

In shift-symmetric Horndeski theories 
with the coupling functions (\ref{G2})-(\ref{G5}), 
the background equations of motion yield
\ba
h' &=& 
[8 G_2 r^2-8 G_{3,X} r^2 h^2  \phi'^2 \phi''-16 G_4 h 
\nonumber \\
& &
-8( 4 G_{4,X} r \phi'' + 2 G_{4,X} \phi'
-4 G_{4,XX} hr \phi'^2 \phi'')h^2 \phi'  \nonumber \\
& &
+8 (3G_{5,X}-G_{5,XX} h \phi'^2)h^3 \phi'^2 \phi''] 
\nonumber \\
& & 
/\{ 4[4r( G_4+2 G_{4,X} 
h \phi'^2 -G_{4,XX} h^2 \phi'^4)  
\nonumber \\
& & 
+h \phi'^3 (G_{3,X} r^2 
- 5 G_{5,X} h 
+G_{5,XX} h^2 \phi'^2)] \}\,,\label{heq} \\
f' &=& 
2 f ( G_2 r^2 + G_{2,X} h r^2 \phi'^2
-2 G_{3,X} h^2 r \phi'^3 - 2 G_4 h
\nonumber \\
& &
- 4 G_{4,X} h^2 \phi'^2 
+2 G_{4,XX} h^3 \phi'^4) \nonumber \\
& &
/\{ h [ 4r(G_4+ 2 G_{4,X} h \phi'^2 
-G_{4,XX} h^2 \phi'^4)  \nonumber \\
& & 
+h \phi'^3 (G_{3,X} r^2 
- 5 G_{5,X} h 
+G_{5,XX} h^2 \phi'^2)] \}\,,\label{feq} \\
J' &=& 0\,,\label{Jeq}
\ea
where $J$ is a scalar-field current given by 
\ba
J &=&\frac{1}{2} 
\sqrt{\frac{h}{f}} \phi'
[ 2 G_{2,X}f r^2 -G_{3,X} r h \phi' (rf' + 4f) 
\nonumber \\
& &
- 4( G_{4,X}-G_{4,XX} h\phi'^2)  (rf' + f) h \nonumber \\
& &
+( 3 G_{5,X}-G_{5,XX} h \phi'^2) f' h^2 \phi' ]\,.
\label{J2}
\ea
One can integrate Eq.~(\ref{Jeq}) to obtain
\be
J={\cal C}\,,
\label{JC}
\ee
where ${\cal C}$ is a constant. 

\subsection{Model~1}

Let us consider the planar BH solutions in Model~1, 
defined by the coupling functions given 
in Eq.~(\ref{example1}).
In this case, the current (\ref{J2}) reduces to
\be
J(r)=
2 \ell^2 h \sqrt{\frac{h}{f}} \frac{\left( r \phi'-1 \right)^2
\left( f'-2f \phi' \right) \left( 1-5 \ell^2 h \phi'^2 \right)} 
{(1+\ell^2 h \phi'^2)^4}.
\label{Jmo1}
\ee
Suppose that there exists at least one horizon 
located at $r=r_h$.
Near $r=r_h$, the metric components 
$f$ and $h$ can be expanded as
\be
f=\sum_{i=1} f_i (r-r_h)^i\,,\qquad 
h=\sum_{i=1} h_i (r-r_h)^i\,,
\ee
where $f_i$, $h_i$, and $r_h$ are constants. 
We further assume that $\phi'(r)$ remains 
finite at $r=r_h>0$. Under these conditions, 
the current (\ref{Jmo1}) must vanish on the horizon, 
and hence ${\cal C}=0$. This means that 
\be
J(r)=0\,,
\label{J0}
\ee
at any distance $r$.
Equation (\ref{J0}) can be satisfied for 
\be
\phi'(r)=\frac{1}{r}\,.
\label{phiso}
\ee
We also have $J=0$ either 
when $\phi' = f'/(2f)$ or when
$\phi'^2 = 1/(5 \ell^2 h)$.
However, since $\phi'$ diverges on the horizon in 
both cases, we do not consider such
possibilities.\footnote{The solution 
${\phi'}^2=1/(5\ell^2 h)$ leads to 
$X={\rm constant}$. Although $X$ remains 
finite everywhere, this solution does not allow 
$X$ to change sign across the horizon ($h=0$).}
Substituting Eq.~(\ref{phiso}) into 
Eqs.~(\ref{heq}) and (\ref{feq}), we obtain 
\ba
\hspace{-0.5cm}
& &
h' =-\frac{\Lambda r^4 + (2 \Lambda \ell^2 + 1)r^2 h 
+ (\Lambda \ell ^2 + 3) \ell^2 h^2}{r^3}\,,
\label{heq1}\\
\hspace{-0.5cm}
& &
\frac{f'}{f}=\frac{h'}{h}\,.
\label{fh1}
\ea
Integrating Eq.~(\ref{fh1}) yields the relation 
$f={\cal C}_0 h$, where ${\cal C}_0$ is a constant. 
Using the freedom of time reparametrization, we set
${\cal C}_0=1$ without loss of generality. 
We also consider a negative cosmological constant
$\Lambda=-3/\ell_{\Lambda}^2<0$, 
where $\ell_{\Lambda}$ corresponds to the AdS radius.
Integrating Eq.~(\ref{heq1}) gives the following 
solution:
\be
f=h=\left( \frac{r^2}{\ell_{\Lambda}^2}
-\frac{2M}{r} \right)
\left( 1-\frac{\ell^2}{\ell_{\Lambda}^2}
+\frac{2M \ell^2}{r^3} 
\right)^{-1}\,,
\label{fhso}
\ee
where $M$ is a constant. 
We require that $M>0$ to ensure the existence 
of the horizon. Then, the horizon is located 
at the distance $r_h = (2M \ell_{\Lambda}^2)^{1/3}$.
In the limit $\ell\to 0$, the metric components 
reduce to $f=h= r^2/\ell_{\Lambda}^2-2M/r$, 
which correspond to the planar BH solution 
in GR on an asymptotically AdS 
background \cite{Lemos:1994fn}.

For $\ell \neq 0$, expanding Eq.~(\ref{fhso}) around $r=0$ yields
\be
f=h=-\frac{r^2}{\ell^2}+\frac{r^5}{2\ell^4 M}+{\cal O}(r^8)\,,
\label{fasy1}
\ee
indicating that $f$ and $h$ are regular at $r=0$.
Around $r=0$, the above planar BHs are spacelike ($f=h<0$), 
unlike regular BHs with de Sitter centers, 
for which the metric components take the form 
$f = h = 1 - \Lambda r^2 + \cdots$ \cite{Bardeen:1968,Ayon-Beato:1998hmi,Dymnikova:2004zc,Hayward:2005gi}.
At large distances, the metric components have 
the following asymptotic behavior 
\be
f=h=\frac{r^2}{\ell_{\Lambda}^2-\ell^2}
-\frac{2M \ell_{\Lambda}^4}{(\ell_{\Lambda}^2-\ell^2)^2r}
+{\cal O}(r^{-3})\,.
\label{fasy2}
\ee
From this expression we can read off an effective 
cosmological constant $\Lambda_{\rm eff}
=-3/(\ell_{\Lambda}^2-\ell^2)$, and an ADM mass 
$\mu=M \ell_{\Lambda}^4/(\ell_{\Lambda}^2-\ell^2)^2$. 
For the BH to be timelike at spatial infinity, we require 
the condition $\ell_{\Lambda}^2>\ell^2>0$, or, equivalently, 
for $-3/\ell^2<\Lambda<0$, the variable $r$ 
is spacelike.\footnote{For the fine-tuned case 
$\Lambda=-3/\ell^2$, the metric component is given by 
$h=r^5[c_1-1/(\ell^2 r^3)]$, where $c_1$ is a positive constant.}

\begin{figure}[ht]
\begin{center}
\includegraphics[height=3.0in,width=3.0in]{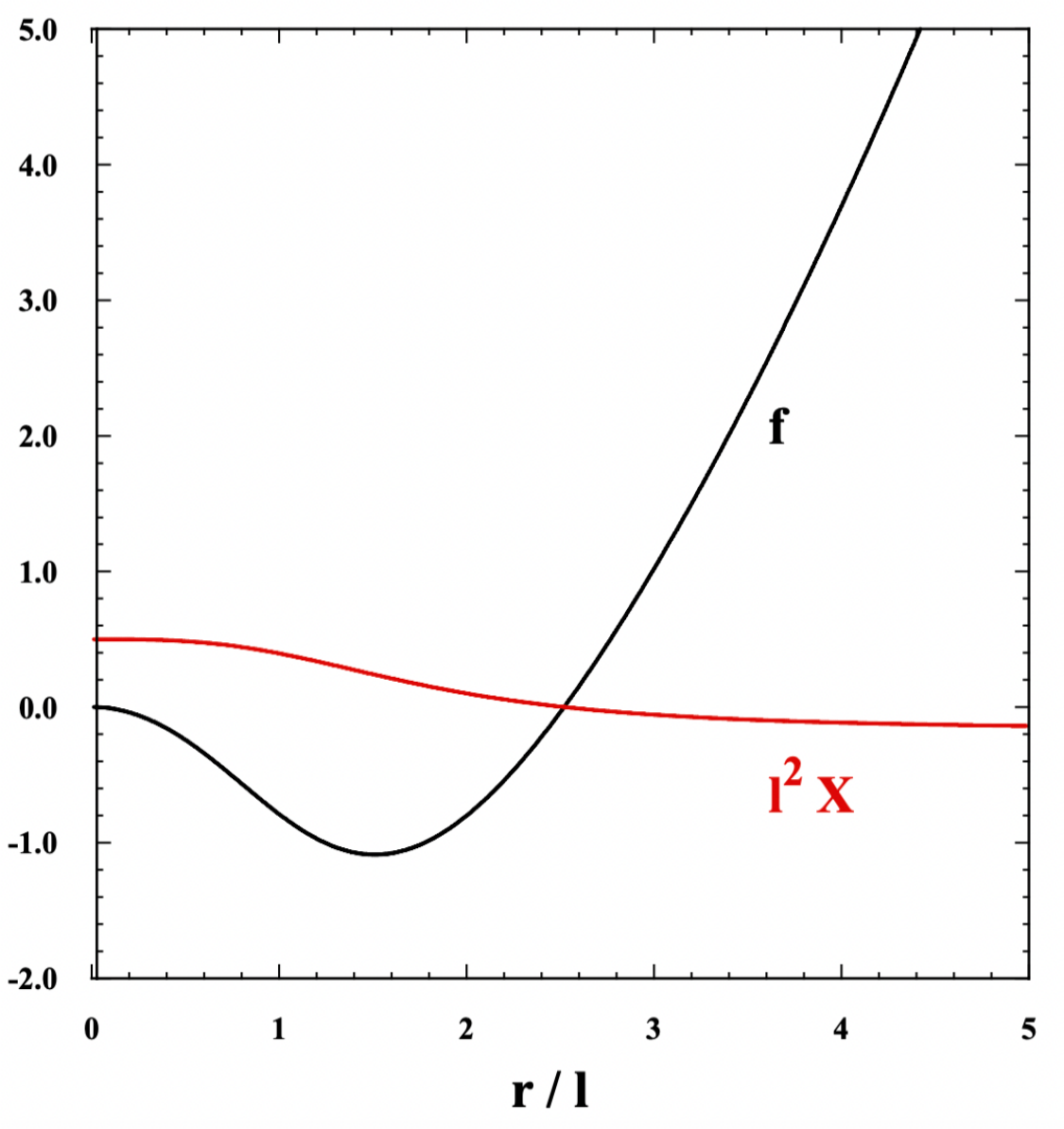}
\end{center}\vspace{-0.5cm}
\caption{
The metric function $f~(=h)$ and the rescaled scalar 
derivative $ \ell^2 X$ are plotted as functions of $r/\ell$ 
in the range $r \geq 0$ for Model~1
with $M=\ell_{\Lambda}$ and  
$\ell=\ell_{\Lambda}/2$. 
The horizon is located at 
$r_h=2^{1/3}\ell_{\Lambda}=2^{4/3}\ell$, where both 
$f$ and $X$ vanish. The kinetic term 
decreases smoothly from the value $X=1/(2\ell^2)$ at $r=0$ 
to the asymptotic value $X=-1/[2(\ell_{\Lambda}^2-\ell^2)]$, 
as $r \to \infty$.
\label{fig1}
}
\end{figure}

In the limits $r \to 0$ and $r \to \infty$, the 
scalar-field kinetic term $X=-(1/2)h \phi'^2$ 
approaches $1/(2\ell^2)$ and 
$-1/[2(\ell_{\Lambda}^2-\ell^2)]$, respectively.
As long as $X$ is a continuously decreasing 
function of $r$, it should be in the range 
\be
-\frac{1}{2(\ell_{\Lambda}^2-\ell^2)}<X
<\frac{1}{2\ell^2}\,.
\ee
Note that $X$ crosses 0 at the horizon ($h=0$). 
In Fig.~\ref{fig1}, we plot $\ell^2 X$ and $f\,(=h)$ 
versus $r/\ell$ in the range $r \geq 0$ for $M=\ell_{\Lambda}$ 
and $\ell=\ell_{\Lambda}/2$.
The quantity $X$ decreases smoothly from 
$1/(2\ell^2)$ at $r=0$ to  
$-1/[2(\ell_{\Lambda}^2-\ell^2)]$ at spatial infinity.  
For $r < r_h$, the metric function is negative ($f < 0$), 
whereas it becomes positive ($f > 0$) for $r > r_h$.
At $r = r_h$, both $X$ and $f$ vanish.
 
%%%%%%%%%%%%%%%%%%%%%%%%%%%%%%%%
%%%%%%%%%%%%%%%%%%%%%%%%%%%%%%%%
 
%%%%%%%%%%%%%%%%%%%%%%%%%%%%%%%%
\begin{figure}[ht]
\begin{center}
\includegraphics[height=3.0in,width=3.0in]{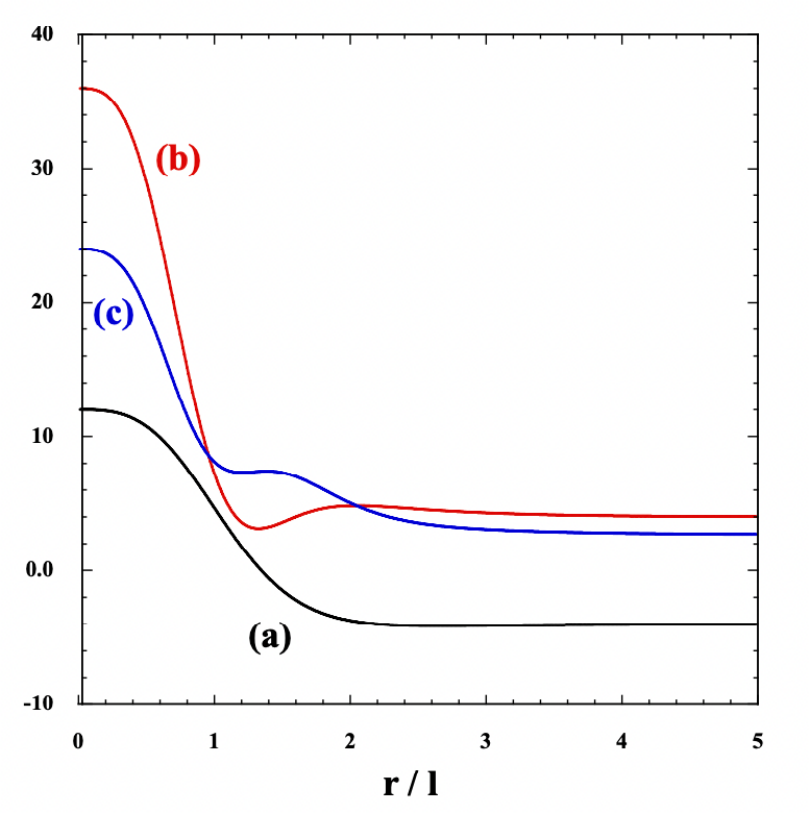}
\end{center}\vspace{-0.5cm}
\caption{
We plot 
(a) $\ell^2 R$, 
(b) $\ell^4 R_{\alpha \beta} R^{\alpha \beta}$, 
and 
(c) $\ell^4 R_{\alpha \beta \mu \nu} R^{\alpha \beta \mu \nu}$
as functions of $r/\ell$ for Model~A with the same values of 
$M$ and $\ell$ as those used in Fig.~\ref{fig1}.
These curvature invariants are regular 
for all $r \geq 0$.
\label{fig2}
}
\end{figure}
%%%%%%%%%%%%%%%%%%%%%%%%%%%%%%%% 
 
Using the solution (\ref{fhso}), we compute the Ricci scalar, the squared Ricci tensor, and the Kretschmann scalar.  
Expanding these quantities around $r=0$, we find 
\ba
R &=& \frac{12}{\ell^2}-\frac{21}{\ell^4 M}r^3+{\cal O} (r^6)\,,\label{cur1}\\
R_{\alpha \beta} R^{\alpha \beta}  &=&
\frac{36}{\ell^4}-\frac{126}{\ell^6 M}r^3+{\cal O} (r^6)\,,\\
R_{\alpha \beta \mu \nu} R^{\alpha \beta \mu \nu}  &=&
\frac{24}{\ell^4}-\frac{84}{\ell^6 M}r^3+{\cal O} (r^6)\,,
\label{cur3}
\ea
all of which remain finite at $r=0$.
As shown in Fig.~\ref{fig2}, these three curvature invariants 
remain finite not only at $r=0$ but also at all values of $r$.
Thus, at least at the background level, the planar BHs discussed above are regular and possess a finite kinetic term $X$ for $r \geq 0$.

The fact that both $f$ and $h$ vanish at $r=0$ suggests that the coordinate point $r=0$ corresponds to an inner horizon. In this case, one may investigate whether the coordinate system can be extended to the region $r<0$.
Indeed, there exists a point 
\be
r=r_{\infty}=-(2M\ell^2)^{1/3}\,(1-\ell^2/\ell_\Lambda^2)^{-1/3}<0\,,
\ee
at which 
the metric component exhibits the divergence: 
$f(r \to +r_{\infty})=-\infty$. 
In the region $r_{\infty}<r<0$, we have $f=h<0$, and hence 
the coordinate $r$ plays the timelike role.
We then define a new coordinate $\rho$ via $r = r_\infty + \rho$, where $\rho > 0$ serves as a coordinate orthogonal to the $x_2$-$x_3$ plane. 
The regularity at $r=0$ (i.e., $\rho = -r_\infty > 0$) may not be sufficient to define a truly regular BH in this extended coordinate chart, since a genuine singularity appears as $\rho \to +0$ in Model~1, where both $|X|$ and $R$ diverge.
In GR, such an extension is not 
possible because there is a physical 
singularity at $r=0$. 
As we will see in Secs.~\ref{oddsec} and \ref{evensec}, the BH solution~(\ref{fhso}) 
with $\phi'(r)=1/r$ suffers from strong coupling and instability problems 
in the region $r \geq 0$. 
Since this does not give any further motivation to explore the properties of the solution beyond the point $r=0$, we will not discuss those details here.

\subsection{Model~2}

In Model~2, which is defined by the coupling functions 
in Eq.~(\ref{example2}), the scalar-field current 
is given by  
\be
J=2\ell^4 h^2 \phi'^2 \sqrt{\frac{h}{f}}
\frac{
\left( r \phi'-1 \right)^2
\left( f'-2f \phi' \right) \left( 3+5 \ell^4 h^2 
\phi'^4 \right)}
{(\ell^4 h^2 \phi'^4-1)^3}.
\ee
The integration constant ${\cal C}$ in Eq.~(\ref{JC}) 
must be 0 to satisfy the boundary condition 
$h = 0$ at the horizon.
Consequently, the solution to $J=0$, which gives
a finite nonvanishing value of $\phi'$ in the range 
$r>0$, is the same as Eq.~(\ref{phiso}).
Substituting $\phi'(r)=1/r$ into Eqs.~(\ref{heq}) 
and (\ref{feq}) yields:
\ba
& &
h'=\frac{2h r^2-\Lambda (r^4-h^2 \ell^4)}{r^3} 
\nonumber \\
& &
\qquad+\frac{3(r^4-h^2 \ell^4)}{2 \ell^2 r^3} 
\ln \left( \frac{r^2-h \ell^2}{r^2+h \ell^2} \right)\,, 
\label{heq2} \\
& & 
\frac{f'}{f}=\frac{h'}{h}\,.
\label{fh2}
\ea
From Eq.~(\ref{fh2}), we obtain the solution 
$f={\cal C}_0 h$, where the constant ${\cal C}_0$ 
is chosen to be 1 in the following. 
Integrating Eq.~(\ref{heq2}) and setting 
$\Lambda=-3/\ell_{\Lambda}^2$, we obtain
\be
f=h=-\frac{r^2}{\ell^2} \tanh (x)\,,
\label{fhmoB}
\ee
where 
\be
x=-\frac{\ell^2}{\ell_{\Lambda}^2}
+\frac{2M \ell^2}{r^3}\,,
\label{xdef}
\ee
and $M$ is an integration constant. 
We will consider the case in which $M$ is positive 
to ensure the existence of the horizon at $x=0$. 
In this case, the horizon is located 
at $r_h=(2M \ell_{\Lambda}^2)^{1/3}$. 
For $r \to +0$, we have $x \to +\infty$ and 
$f=h \to -r^2/\ell^2<0$, so the metric remains regular 
with $r$ acting as a timelike coordinate. 
At spatial infinity ($r \to \infty$), $f$ and $h$ have the 
dependence $f=h \simeq (r^2/\ell^2)\tanh 
(\ell^2/\ell_{\Lambda}^2)$. 
Using the solution (\ref{fhmoB}), the scalar-field 
kinetic term is given by
\be
X=\frac{1}{2\ell^2} \tanh (x)\,, 
\ee
which approaches the finite value $X \to 1/(2\ell^2)$ 
for $r \to +0$. In the other limit $r \to \infty$, 
$X$ approaches $-(1/2\ell^2)\tanh (\ell^2/\ell_{\Lambda}^2)$. 
Therefore, $X$ is bounded in the range 
\be
-\frac{1}{2\ell^2} \tanh \left( \frac{\ell^2}
{\ell_{\Lambda}^2} \right)< X < \frac{1}{2\ell^2}\,.
\ee
At $r=r_h$, we have $X=0$.
These features of $f$, $h$, and $X$ are similar 
to those discussed in Model~1. 
Indeed, the leading-order contributions to $R$, $R_{\alpha \beta} R^{\alpha \beta}$, and $R_{\alpha \beta \mu \nu} R^{\alpha \beta \mu \nu}$ coincide with those in Eqs.~(\ref{cur1})-(\ref{cur3}).
Therefore, as in Model~1, the curvature invariants 
remain finite at $r=0$. 

Note that, in the limit $r \to -0$, we have 
$f=h \simeq r^2/\ell^2 \to +0$, which corresponds to 
the region where $r$ plays the role of 
a spacelike coordinate. Thus, the metric components 
change sign across the point $r=0$. 
In contrast to Model~1, the divergences of 
$f$ and $h$ do not occur in the region $r<0$.

\subsection{Models corresponding to each value of $n$}

We also consider models in which 
$G_{2,3,4,5}(X)$ correspond to the terms for each $n$; 
for example, $G_3(X) = (2\ell^2)^{-1}
c_n \ell^{2n} G_{3}^{(n)}(X)$ for each $n$.
In this case, the scalar-field current 
takes the following form:
\ba
J &=& 
(-1)^n n (n-1)(2n-3)c_n 
\ell^{2(n-1)} h^{n-1} 
\sqrt{\frac{h}{f}}
\phi'^{2(n-2)} \nonumber\\
& & \times
(r\phi'-1)^2(f'-2f \phi')\,,
\ea
which holds for $n \geq 2$. 
The boundary condition at the horizon, which requires a finite value of $\phi'$, implies that the integration constant 
${\cal C}$ in Eq.~(\ref{JC}) must vanish, 
i.e., ${\cal C}=0$. Consequently, we have $J=0$. 
The solution to $J=0$ that yields a nonvanishing finite value of $\phi'(r)$ for $r>0$ corresponds to $\phi'(r) = 1/r$.
Substituting this solution into Eqs.~(\ref{heq}) 
and (\ref{feq}) gives the relation $f'/f=h'/h$, 
in addition to a differential equation for $h$. 
With an appropriate time parametrization, 
it then follows that $f=h$.
The radial dependence of $f$ and $h$ can be obtained 
by numerically integrating the differential equation for $h$.

The above features are the same as those 
discussed for Models 1 and 2.
The properties $J=0$ and the existence of the solution 
$\phi'(r)=1/r$ hold regardless of the choice of 
coefficients $c_n$.
By taking the infinite sum over $n$, the metric components and curvature invariants can be regular at $r=0$, 
as in Models 1 and 2.
However, this regularity does not generally hold for models with a fixed power of $n$.
 
%%%%%%%%%%%%%%%%%%%%%%%%%%%%%%%%%%%%%%%%%%%%
\section{Perturbations around the planar BH} 
\label{persec}
%%%%%%%%%%%%%%%%%%%%%%%%%%%%%%%%%%%%%%%%%%%%

The linear stability of planar BHs can be 
analyzed by considering perturbations 
on the background (\ref{line}).
We write the metric tensor in the form 
$g_{\mu \nu}=\bar{g}_{\mu \nu}+h_{\mu \nu}$, 
where $\bar{g}_{\mu \nu}$ is the background 
value and $h_{\mu \nu}$ is the metric perturbation. 
In general, the components of 
$h_{\mu \nu}$ are given by 
\ba
& &
h_{tt}=f(r) H_0 (t,r) 
Z(x_2,x_3), \quad 
h_{tr}=H_1 (t,r) Z(x_2,x_3),
\nonumber \\
& &
h_{t a}=h_0(t,r)
\partial_{a} Z(x_2,x_3)+Q(t,r)
\varepsilon_{ab} 
\delta^{bc} \partial_c Z(x_2,x_3)\,,
\nonumber \\
&& 
h_{rr}=h^{-1}(r) H_2(t,r) 
Z(x_2,x_3),\nonumber \\
\hspace{-0.1cm}
& &
h_{ra}=h_1 (t,r) 
\partial_{a}Z(x_2,x_3)
+W(t,r) \varepsilon_{ab} 
\delta^{bc} \partial_c Z(x_2,x_3)\,,
\nonumber \\
\hspace{-0.1cm}
& &
h_{ab}=r^2 K(t,r) \delta_{ab} 
Z(x_2,x_3)+r^2 G(t,r) 
\partial_a \partial_b Z(x_2,x_3) 
\nonumber\\
& &
+\frac{1}{2}U(t,r) 
[{\varepsilon_a}^c 
\partial_c \partial_b Z(x_2,x_3)
+{\varepsilon_b}^c 
\partial_c \partial_a 
Z(x_2,x_3)],
\label{hcom}
\ea
where $H_0$, $H_1$, $h_0$, $Q$, $H_2$, 
$h_1$, $W$, $K$, $G$, $U$ are functions 
of $t$ and $r$, $Z$ is a function of $x_2$ 
and $x_3$, the subscripts $a$, $b$, $c$  
denote either $x_2$ or $x_3$, 
$\partial_a Z \equiv \partial Z/\partial a$, 
and $\varepsilon_{ab}$ is an 
antisymmetric symbol with 
$\varepsilon_{x_2 x_3}=1$.
Since we are considering the two-dimensional flat space with coordinates $(x_2, x_3)$, we can use partial derivatives instead of covariant derivatives---for example, writing $\partial_a \partial_b Z$ in place of $\nabla_a \nabla_b Z$.
We also decompose the scalar field 
$\phi$, as 
\ba
\phi = 
\bar{\phi}(r)+\delta \phi (t,r) 
Z(x_2,x_3)\,,
\label{perma}
\ea
where $\bar{\phi}(r)$ denotes the background component, 
while the perturbed part is given by the product 
of $\delta \phi(t, r)$ and $Z(x_2,x_3)$.
For simplicity, we omit the overbar in what follows.

The perturbed fields $H_0$, $H_1$, $h_0$, $H_2$, $h_1$, $K$, $G$, and $\delta \phi$ correspond to those in the even-parity sector, 
while $Q$, $W$, $U$ are those in the 
odd-parity sector.
The function $Z$ satisfies the relation 
\be
\delta^{ab}\partial_a\partial_b Z
+k^2 Z=0\,,
\ee
where $k$ is the eigenvalue corresponding to 
the wavenumber associated with the two-dimensional 
coordinates $(x_2, x_3)$. 
Namely, $Z$ obeys the differential 
equations, $\partial_{x_2}^2 Z+k^2 Z=0$ 
and $\partial_{x_3}^2 Z+k^2 Z=0$. 
The corresponding solutions are given by 
$Z=A\cos(kx_2+\theta_A)$ and 
$Z=B\cos(kx_3+\theta_B)$,  
where $A$, $\theta_A$, $B$, and $\theta_B$ 
are constants. Under a suitable rotation 
in the $(x_2,x_3)$ plane, we can consider the 
case in which $Z$ is a function of 
$x_2$ alone, i.e., $Z=A\cos(kx_2+\theta_A)$. 
Setting $\theta_A=0$ without loss of 
generality and imposing 
\be
\int_{-\pi}^{\pi} Z^2\, {\rm d}x_2=1\,,
\label{nor}
\ee
for an integer $k$, the mode function 
satisfying this normalization condition 
takes the form 
\be
Z=\frac{1}{\sqrt{\pi}} \cos (k x_2)\,.
\label{eigen}
\ee
We will use this form of $Z$ together with 
the normalization condition (\ref{nor}) 
for deriving the second-order action 
of perturbations. We have restricted the range 
of $x_2$ to $-\pi \le x_2 \le \pi$, as in the cases of 
cylindrical and toroidal BHs. 
However, the local stability conditions remain 
the same as those for planar BHs.
We will further consider a finite range of $x_3$, 
for instance, $0 \le x_3 \le 1$, without loss of generality.
We note that the mode function 
here is simpler than the spherical harmonics 
$Y_{lm}(\theta, \varphi)$ used in the analysis 
for BH perturbations on the 
SSS background \cite{Regge:1957td,Zerilli:1970se}.

There are residual gauge degrees of freedom 
associated with the coordinate 
transformation, $\tilde{x}^{\mu}=x^\mu+\xi^\mu$. 
In general, the change of metric 
perturbations $h_{\mu \nu}$ under this transformation 
is given by
\be
\Delta h_{\mu\nu}=-g_{\mu\alpha}
\xi^\alpha{}_{,\nu}-g_{\nu\alpha}
\xi^\alpha{}_{,\mu}-g_{\mu\nu,\alpha}
\xi^\alpha\,,
\ee
where $\xi^\alpha{}_{,\nu}=\partial 
\xi^{\alpha}/\partial x^{\nu}$, 
$g_{\mu\nu,\alpha}=\partial g_{\mu\nu}/
\partial x^{\alpha}$, and 
the components of $\xi^\alpha$ are 
\be
\xi^\alpha=(\xi^t,\xi^r,\xi^a)^T\,,\quad 
{\rm with} \quad 
\xi^a=\delta^{ac}\partial_c\xi_S+\xi^a_V\,.
\label{xia}
\ee
Note that $\xi_S$ is a scalar mode, and $\xi_V^a$ 
is a vector mode with respect to 
two-dimensional rotations in the $(x_2, x_3)$ 
plane, 
satisfying the transverse condition 
$\partial_a \xi_V^a = 0$.
In terms of a pseudo scalar field $\xi_V$, 
we can express the covariant component of 
$\xi_V^a$ as $(\xi_V)_a
\equiv \delta_{ab}\xi^b_V
=\varepsilon_{ac}\delta^{cd}\partial_d \xi_V$. 
Each variable is further decomposed 
in terms of the eigenfunction $Z$ of 
the two-dimensional Laplacian. 
For instance, we have $\xi_V=\xi_V(t,r)\,Z(x_2,x_3)$, 
and hence, 
\be
\xi_V^a=\delta^{ab}\varepsilon_{bc}
\delta^{cd}\xi_V(t,r)\partial_d Z\,.
\ee

The change of the $(x_2, x_3)$ components of 
metric perturbations under the gauge 
transformation is given by 
\ba
\Delta h_{ab} &=&
-2r\xi^rZ\delta_{ab}
-2r^2\xi_S\partial_a\partial_bZ 
\nonumber \\
& &
-r^2\xi_V \left( \varepsilon_a{}^d\partial_d
\partial_bZ+\varepsilon_b{}^d
\partial_d\partial_aZ \right)\,,
\ea
where $\varepsilon_a{}^d=\varepsilon_{ab}
\delta^{bd}$. Then, the perturbed fields 
$K$, $G$, $U$, $Q$, and $W$ transform, 
respectively, as
\ba
& &
\Delta K=-\frac{2}{r}\xi^r\,,\quad 
\Delta G =-2\xi_S\,,\quad 
\Delta U=-2r^2\xi_V\,,\nonumber \\
& &
\Delta Q=r^2\dot{\xi}_V\,,\quad 
\Delta W=r^2\xi_V'\,,
\ea
where a dot represents the derivative 
with respect to $t$. 
The transformations of $h_0$, $h_1$, $H_0$, 
$H_1$, and $H_2$ are given, respectively, as 
\ba
& &
\Delta h_0 = 
f\xi^t-r^2\dot{\xi}_S\,,\quad
\Delta h_1 = 
-\frac{\xi^r}{r}-r^2\xi_S'\,,\nonumber \\
& &
\Delta H_0 = 2f\dot{\xi}^t+f'\xi^r\,,\quad
\Delta H_1 = f\dot{\xi}^t-\frac{\dot{\xi}^r}{h}\,,
\nonumber \\
& &
\Delta H_2 = \frac{h'}{h}\xi^r-2(\xi^r)'\,.
\ea
In the following, we impose the gauge conditions
\be
K=0\,,\qquad G=0\,,\qquad 
h_0=0\,,\qquad U=0\,,
\label{gaugecon}
\ee
under which the components $\xi^t$, $\xi^r$, 
$\xi_S$, and $\xi_V$ are completely fixed. 
Note that this choice is analogous to the gauge 
conditions adopted in Refs.~\cite{Kobayashi:2012kh,Kobayashi:2014wsa,Kase:2021mix,Kase:2023kvq} within the context of Horndeski theories.

%%%%%%%%%%%%%%%%%%%%%%%%%%%%%
\section{Odd-parity perturbations} 
\label{oddsec}
%%%%%%%%%%%%%%%%%%%%%%%%%%%%%

In this section, we derive the stability 
conditions for odd-parity perturbations 
and apply them to the models discussed 
in Sec.~\ref{BHsec}. 
For the choice of the eigenfunction 
(\ref{eigen}), the nonvanishing components 
of perturbations in the odd-parity 
sector are given by
\ba
h_{t x_3} &=&
Q(t,r) \frac{k}{\sqrt{\pi}} 
\sin (k x_2)\,,\\
h_{r x_3} &=&
W(t,r) \frac{k}{\sqrt{\pi}} 
\sin (k x_2)\,.
\ea
In shift-symmetric Horndeski theories with 
the coupling functions $G_{2,3,4,5}(X)$, 
we expand the action up to second order 
in odd-parity perturbations. 
Using the normalization condition (\ref{nor}) 
for the integration over $x_2$, 
integrating over the range 
$0 \le x_3 \le 1$, 
and performing integration by parts, 
the resulting quadratic-order action 
takes the form
\be
{\cal S}_{\rm odd}=k^2 \int {\rm d}t 
{\rm d}r\,{\cal L}_{\rm odd}\,,
\ee
where 
\ba
{\cal L}_{\rm odd}
&=& \frac{1}{4}
\sqrt{\frac{h}{f}} {\cal H}
\left( \dot{W}-Q'+\frac{2Q}{r} 
\right)^2 \nonumber \\
& &+\frac{k^2}{4r^2}\sqrt{\frac{h}{f}} 
\left( \frac{{\cal F}}{h}Q^2
-f {\cal G} W^2 \right)\,,
\label{Lodd}
\ea
with 
\ba
{\cal H} &=& 2G_4+2h \phi'^2 G_{4,X}
-\frac{h^2 \phi'^3 G_{5,X}}{r}\,,\\
{\cal F} &=& 2G_4-h\phi'^2 \left( 
\frac{1}{2}h' \phi'+h \phi'' \right)G_{5,X}\,,\\
{\cal G} &=& 2G_4+2h \phi'^2 G_{4,X}
-\frac{f'h^2 \phi'^3 G_{5,X}}{2f}\,.
\ea
The Lagrangian (\ref{Lodd}) is valid for both 
the regions $f>0$, $h>0$ and $f<0$, $h<0$.

To identify the dynamical degree of freedom, 
we consider the Lagrangian in the form 
\ba
\tilde{{\cal L}}_{\rm odd} &=& 
\frac{1}{4}
\sqrt{\frac{h}{f}} {\cal H} 
\left[ 2\chi \left( \dot{W}-Q'
+\frac{2Q}{r} \right)-\chi^2
\right] \nonumber \\
& &+\frac{k^2}{4r^2}\sqrt{\frac{h}{f}} 
\left( \frac{{\cal F}}{h}Q^2
-f {\cal G} W^2 \right)\,.
\label{Lodd2}
\ea
Varying $\tilde{{\cal L}}_{\rm odd}$ with 
respect to the Lagrange multiplier $\chi$ yields
\be
\chi=\dot{W}-Q'+\frac{2Q}{r}\,.
\label{chi}
\ee
Substituting Eq.~(\ref{chi}) into Eq.~(\ref{Lodd2}), 
we find that $\tilde{{\cal L}}^{\rm odd}$ is equivalent to ${\cal L}^{\rm odd}$. 
Varying Eq.~(\ref{Lodd2}) with respect to $W$ and 
$Q$, we obtain
\ba
W &=& -\frac{r^2 {\cal H}}
{k^2 f {\cal G}} \dot{\chi}\,,\label{W}\\
Q &=& -\frac{r[ 2r fh (\chi {\cal H}' 
+ \chi' {\cal H})
+( rf h'-r f' h + 4fh) \chi {\cal H}]}
{2k^2 f{\cal F}}. \label{Q} \nonumber \\
\ea
Substituting Eqs.~(\ref{W}), (\ref{Q}), and 
their $t$ and $r$ derivatives into 
Eq.~(\ref{Lodd2}) and performing the integration 
by parts, the second-order Lagrangian reduces to  
\be
\tilde{{\cal L}}_{\rm odd}=K_{\chi} \dot{\chi}^2
+G_{\chi} \chi'^2+M_{\chi} \chi^2\,,
\label{Loddf}
\ee
where 
\ba
K_{\chi} &=& \frac{r^2 {\cal H}^2}
{4k^2{\cal G}} \frac{h}{f^2} 
\sqrt{\frac{f}{h}}\,,\\
G_{\chi} &=& -fh \frac{{\cal G}}
{{\cal F}} K_{\chi}\,,\\
M_{\chi} &=& -\frac{f {\cal G}}
{r^2 {\cal H}}K_{\chi} 
\left( k^2+\alpha_{\chi}'-\frac{2}{r}\alpha_{\chi}
\right)\,,
\ea
where 
\be
\alpha_{\chi}= 
-\frac{r^2 h {\cal H}}{{\cal F}} 
\left( \frac{{\cal H}'}{{\cal H}}
-\frac{f'}{2f}+\frac{h'}{2h}
+\frac{2}{r} \right)\,.
\ee
The reduced Lagangian (\ref{Loddf}) shows that 
there is a single dynamical gravitational 
perturbation, $\chi$, in the odd-parity sector.

\subsection{Stability conditions}

\subsubsection{Region with $f>0$ and $h>0$}

In the region characterized by $f>0$ and $h>0$,  
the coordinate $t$ plays a timelike role. 
The absence of ghosts requires that 
$K_{\chi} > 0$, which translates 
into the condition
\be
{\cal G}>0\,.
\label{G}
\ee
To obtain the propagation speed in the radial direction, 
we vary Eq.~(\ref{Loddf}) with respect to 
$\chi$ and substitute the WKB-form solution 
$\chi=\chi_0 e^{-i(\omega t-k_r r)}$ into the 
perturbation equation for $\chi$, where 
$\chi_0$, $\omega$, and $k_r$ are constants. 
In the large $\omega$ and $k_r$ limits, we obtain 
the dispersion relation 
$\omega^2 K_{\chi}+k_r^2 G_{\chi} \simeq 0$. 
The radial propagation speed, which is 
defined as $c_{r,{\rm odd}}=h^{-1/2}
{\rm d}r/{\rm d}\tau$ in proper time 
$\tau=\int \sqrt{f}\,{\rm d}t$, can be 
obtained by substituting $\omega=\sqrt{fh}\,c_r k_r$ 
into the dispersion relation. 
This yields 
\be
c_{r,{\rm odd}}^2=-\frac{G_{\chi}}{fh K_{\chi}}
=\frac{{\cal G}}{{\cal F}}\,.
\label{crodd1}
\ee
The angular propagation speed measured by the
proper time is given by
$c_{\Omega,{\rm odd}}=r{\rm d}l/{\rm d}\tau
=(r/\sqrt{f})(\omega/k)$. 
In the limits of large $\omega$ and $k$, 
the dispersion relation yields 
$\omega^2 K_{\chi}+M_{\chi} \simeq 0$, 
with $M_{\chi} \simeq -f{\cal G}K_{\chi}k^2/
(r^2 {\cal H})$.
Then, we obtain 
the squared angular propagation speed, as 
\be
c_{\Omega,{\rm odd}}^2=-\frac{r^2}{f}
\frac{M_{\chi}}{k^2 K_{\chi}}
=\frac{{\cal G}}{{\cal H}}\,.
\label{cOodd1}
\ee
To avoid the Laplacian instability of 
odd-parity perturbations, 
we require that 
\be
{\cal F}>0\,,\qquad 
{\cal H}>0\,,
\label{FH}
\ee
along with the ghost-free condition (\ref{G}).
We note that these stability conditions are 
the same as those derived for BHs 
on the SSS background \cite{Kobayashi:2012kh}.

\subsubsection{Region with $f<0$ and $h<0$}

In the region characterized by $f<0$ and $h<0$, 
the coordinate $r$ plays a timelike role. 
In this regime, the no-ghost condition 
corresponds to $G_{\chi}>0$, 
which translates to
\be
{\cal F}>0\,.
\ee
To obtain the propagation speeds of 
odd-parity perturbations, we substitute 
a solution of the form 
$\chi=\chi_0 e^{-i(\omega r-k_r t)}$ into 
the perturbation equation for $\chi$.
In the limits of large $\omega$ and $k_r$, 
the dispersion relation yields 
$k_r^2 K_{\chi}+\omega^2 G_{\chi} 
\simeq 0$. The radial propagation speed 
$c_{r,{\rm odd}}=
\sqrt{-f}\,{\rm d}t/{\rm d} \hat{r}$
measured in 
proper time $\hat{r}=\int (-h)^{-1/2}{\rm d}r$ 
is expressed as $c_{r,{\rm odd}}
=\sqrt{fh}\,{\rm d}t/{\rm d}r
=\sqrt{fh}\,\omega/k_r$. 
Then, we obtain
\be
c_{r,{\rm odd}}^2
=-fh \frac{K_{\chi}}{G_{\chi}}
=\frac{{\cal F}}{{\cal G}}\,,
\ee
whose expression is the inverse of  
Eq.~(\ref{crodd1}).
In the limits of large $\omega$ and $k$, 
we approximately obtain the dispersion relation 
$\omega^2 G_{\chi}+M_{\chi} \simeq 0$, 
with $M_{\chi} \simeq (k^2/r^2 h)
({\cal F}/{\cal H})G_{\chi}$. 
Since the angular propagation speed 
is given by 
$c_{\Omega,{\rm odd}}=
r {\rm d} l/{\rm d}\hat{r}
=r\sqrt{-h}\, \omega/k$, we have 
\be
c_{\Omega,{\rm odd}}^2=\frac{r^2 h}{k^2} 
\frac{M_{\chi}}{G_{\chi}}
=\frac{{\cal F}}{{\cal H}}\,,
\ee
whose numerator differs from 
that in Eq.~(\ref{cOodd1}).

If the three conditions ${\cal F} > 0$, ${\cal G} > 0$, 
and ${\cal H} > 0$ are satisfied, the odd-parity 
perturbation is free from both ghost and Laplacian instabilities. 
This stability criterion coincides with that in the regime 
where $f>0$ and $h>0$.

\subsection{Application to concrete models}

Let us examine the linear stability of regular 
planar BHs discussed in Sec.~\ref{BHsec}. 
In all these models, we have shown the existence of a background solution characterized by $\phi'(r)=1/r$. For this solution, the metric functions satisfy $f=h$. Using these properties, 
the difference between ${\cal F}$ and ${\cal G}$ 
is given by 
\be
{\cal F}-{\cal G}=-\frac{h}{r^4} 
\left( 2 r^2 G_{4,X}-h\,G_{5,X} 
\right)\,.
\ee
The functions $G_4^{(n)}$ and $G_5^{(n)}$, 
which are given in Eq.~(\ref{Gchoice}), satisfy 
$2r^2 G_{4,X}^{(n)}=hG_{5,X}^{(n)}$. 
Then, from Eqs.~(\ref{G4}) and (\ref{G5}),  
we obtain the following relation:
\be
2 r^2 G_{4,X} = h G_{5,X}\,, 
\ee
irrespective of the choice of coefficients $c_n$.
Then, for the solution $\phi'(r)=1/r$ with 
$f=h$, we generally have the following property:
\be
{\cal F}={\cal G}
=2G_4-\frac{G_{4,X}}{r^2}(rh'-2h)\,.
\ee
This leads to
\be
c_{r, {\rm odd}}^2 = 1\,,
\ee
at any distance $r$, and hence 
the radial propagation speed 
is always luminal.
The squared angular propagation 
speed takes the form 
\be
c_{\Omega,{\rm odd}}^2=\frac{{\cal G}}{{\cal H}}
=\frac{{\cal F}}{{\cal H}}
=1-\frac{G_{4,X}}{2G_4 r^2}(rh'-2h)
\,,
\ee
irrespective of the sign of $f$. 
Since the values of ${\cal F}~(={\cal G})$ and 
$c_{\Omega, {\rm odd}}^2$ depend on the 
form of the Horndeski functions, 
we explicitly compute them for Models~1 and 2.

\subsubsection{Model~1}

In Model~1, the metric component $h$ 
is given by Eq.~(\ref{fhso}). 
Using this solution, we obtain 
\ba
& &
{\cal F}={\cal G}
=-[10M \ell^2 \ell_{\Lambda}^2 
-(\ell_{\Lambda}^2-\ell^2)r^3] \nonumber \\
&& \qquad \qquad\,
\times 
[2M \ell^2 \ell_{\Lambda}^2 
+(\ell_{\Lambda}^2-\ell^2)r^3]
/(\ell_{\Lambda}^4 r^6)\,,\\
& &
c_{\Omega,{\rm odd}}^2=-\frac{10M \ell^2 \ell_{\Lambda}^2
-(\ell_{\Lambda}^2-\ell^2)r^3}
{2M \ell^2 \ell_{\Lambda}^2
+(\ell_{\Lambda}^2-\ell^2)r^3}\,.
\ea
Expanding these expressions around $r=0$, 
we find
\ba
\hspace{-0.5cm}
& &
{\cal F}={\cal G}
=-\frac{20 M^2 \ell^4}{r^6}
-\frac{8 M \ell^2 
(\ell_{\Lambda}^2-\ell^2)}{\ell_{\Lambda}^2 r^3}+{\cal O}(r^{0})\,,\\
\hspace{-0.5cm}
& &
c_{\Omega,{\rm odd}}^2=-5
+\frac{3(\ell_{\Lambda}^2-\ell^2)}
{M \ell^2 \ell_{\Lambda}^2}r^3
+{\cal O}(r^6)\,.
\ea
Since the leading-order terms of 
${\cal F}$ and $c_{\Omega, {\rm odd}}^2$
are negative, both ghost and Laplacian 
instabilities arise near $r=0$.
This shows that the BH solution (\ref{fhso}) 
with $\phi'(r)=1/r$ is unstable due to the 
rapid growth of odd-parity perturbations around $r=0$.

\subsubsection{Model~2}

In Model~2, we have 
\ba
& &
{\cal F}={\cal G}
=-\frac{\cosh(x)}{r^3} \nonumber \\
& & \qquad \qquad \times 
\left[ 12M \ell^2 \sinh(x) 
-r^3 \cosh(x) \right]\,,
\label{F2} \\
& &
c_{\Omega,{\rm odd}}^2=
1-\frac{12M \ell^2}{r^3} 
\tanh(x)\,,
\label{cO2}
\ea
where $x$ is defined in Eq.~(\ref{xdef}). 
Since $M>0$, we have $x \to +\infty$ 
in the limit $r \to +0$. 
In this limit, Eqs.~(\ref{F2}) and (\ref{cO2}) 
approximately behave as
\be
{\cal F}={\cal G}
\simeq -\frac{3M \ell^2 e^{2x}}{r^3}\,,\qquad 
c_{\Omega,{\rm odd}}^2 \simeq
-\frac{12M \ell^2}{r^3}\,,
\ee
which are both negative.
Therefore, the BH solution (\ref{fhmoB}) is 
unstable due to the presence of both ghost 
and Laplacian instabilities 
near $r=0$. The difference from Model~1 is that 
$c_{\Omega,{\rm odd}}^2$ diverges to 
$-\infty$ in the limit $r \to +0$, 
indicating that the Laplacian instability is 
even more severe in Model~2. 

In both Models~1 and 2, we have shown that planar 
BHs regular at $r=0$ are ruled out due to the 
instability of odd-parity perturbations.
In contrast, in GR, we have 
$G_4 =1/2$ and $G_5=0$, 
which leads to ${\cal F} = {\cal G} 
= 1$ and 
$c_{\Omega, {\rm odd}}^2 = 1$.
In models involving an infinite sum over 
$n$, the function $G_4(X)$ exhibits a nontrivial 
dependence on $X$, as seen in Eqs.~(\ref{example1}) and (\ref{example2}). 
This specific feature leads to 
negative values of ${\cal F}$ 
and $c_{\Omega,{\rm odd}}^2$ near $r=0$.

%%%%%%%%%%%%%%%%%%%%%%%%%%%%%
\section{Even-parity perturbations} 
\label{evensec}
%%%%%%%%%%%%%%%%%%%%%%%%%%%%%

In the even-parity sector, there are five perturbed 
fields $H_0$, $H_1$, $H_2$, $h_1$, and $\delta \phi$ 
by choosing the gauge conditions (\ref{gaugecon}). 
For the function $Z$ given in Eq.~(\ref{eigen}), 
the nonvanishing components of metric perturbations 
are $h_{tt}$, $h_{tr}~(=h_{rt})$, $h_{rr}$, 
and $h_{r \theta}~(=h_{\theta r})$. 
We expand the action up to second-order in 
even-parity perturbations and perform the integrals with respect to $x_2$ and $x_3$. 
After the integration by parts, the resulting 
second-order action is written in the form 
\be
{\cal S}_{\rm even}=\int {\rm d}t 
{\rm d}r\,{\cal L}_{\rm even}\,,
\ee
where 
\begin{widetext}
\ba
{\cal L}_{\rm even} &=& 
H_0 \left[ a_1 \delta \phi'' 
+ a_2 \delta \phi' + a_3 H_2' + k^2 a_4 h_1' 
+ k^2 a_5 \delta \phi + (k^2 a_6 + a_7) H_2 
+ k^2 a_8 h_1 \right] \nonumber \\
& &
+ k^2 b_1 H_1^2 
+ H_1 ( b_2 \dot{\delta \phi}' + b_3 \dot{\delta \phi} 
+ b_4 \dot{H}_2+k^2 b_5 \dot{h}_1) 
+ c_1 \dot{\delta \phi} \dot{H}_2 
+ H_2 ( c_2 \delta \phi'+k^2 c_3 \delta \phi
+k^2 c_4 h_1) 
+ c_5 H_2^2 \nonumber \\
& &
+ k^2 d_1 \dot{h}_1^2 + k^2 h_1 (d_2 \delta \phi' 
+ d_3 \delta \phi) 
+ e_1 \dot{\delta \phi}^2 + e_2 \delta \phi'^2 
+ k ^2 e_3 \delta \phi^2\,.
\label{Lageven}
\ea
\end{widetext}
The $r$-dependent coefficients $a_1$, etc., 
are given in Appendix~B.

Varying the Lagrangian (\ref{Lageven}) 
with respect to $H_0$ and $H_1$, respectively, it follows that 
\ba
& &
a_1 \delta \phi'' 
+ a_2 \delta \phi' + a_3 H_2' + k^2 a_4 h_1' 
+ k^2 a_5 \delta \phi \nonumber \\
& &
+ (k^2 a_6 + a_7) H_2 
+ k^2 a_8 h_1=0\,,\label{H0eq}\\
& &
H_1=-\frac{b_2 \dot{\delta \phi}'
+b_3 \dot{\delta \phi}+b_4 \dot{H}_2
+k^2 b_5 \dot{h}_1}{2k^2 b_1}\,,
\label{H1eq}
\ea
where the latter equation is valid 
for $b_1 \neq 0$.
By using Eqs.~(\ref{H0eq}) and (\ref{H1eq}), 
we can eliminate the fields 
$H_0$ and $H_1$ from Eq.~(\ref{Lageven}).
As a next step, we introduce a dynamical perturbation defined by 
\be
\psi \equiv H_2+\frac{a_1}{a_3} \delta \phi'
+\frac{a_4}{a_3}k^2 h_1\,,
\ee
which allows one to express $H_2$ and $H_2'$ 
in terms of $\psi$, $\delta \phi$, $h_1$, 
and their $r$ derivatives. 
Then, we can eliminate $H_2$ and $H_2'$ 
from Eq.~(\ref{H0eq}). After this procedure, 
the field $h_1$ is expressed in terms of 
$\psi$, $\delta \phi$, and 
their first $r$ derivatives. Using these relations to remove $H_2$ and $h_1$ from Eq.~(\ref{Lageven}) and performing 
the integration by parts, the resulting 
second-order Lagrangian takes the form 
\be
{\cal L}_{\rm even}  = 
\dot{\vec{\mathcal{X}}}^{t}{\bm K} \dot{\vec{\mathcal{X}}}
+\vec{\mathcal{X}}'^{t}{\bm G} \vec{\mathcal{X}}'
+\vec{\mathcal{X}}^{t}{\bm Q} \vec{\mathcal{X}}'
+\vec{\mathcal{X}}^{t}{\bm M} \vec{\mathcal{X}}\,,
\label{evenact}
\ee
where ${\bm K}$, ${\bm G}$, ${\bm M}$ are 
the $2 \times 2$ symmetric matrices, while 
${\bm Q}$ is antisymmetric, 
and the vector $\vec{\mathcal{X}}$ 
is defined as 
\be
\vec{\mathcal{X}}^{t}=\left( \psi, \delta \phi 
\right)\,.
\ee
The perturbations $\psi$ and $\delta \phi$ correspond to the dynamical perturbations originating from the gravitational and scalar-field sectors, respectively.
To derive the no-ghost conditions and radial propagation 
speeds for $\psi$ and $\delta \phi$, 
we introduce the following combinations 
\ba
{\cal P} &\equiv& \frac{h \mu}{2fr^2 {\cal H}^2} 
\left( \frac{fr^4 {\cal H}^4}{\mu^2 h} \right)'\,,\\
\mu &\equiv& \frac{2(\phi' a_1+2r a_4)}{f} 
\sqrt{\frac{f}{h}}\,.
\ea
For the derivation of linear stability conditions, we use the relations among the coefficients given in Appendix~B, e.g., $a_2=a_1'+J/2$. 
Since the scalar-field current $J$ obeys $J'=0$ 
in shift-symmetric Horndeski theories, 
we have $a_2'=a_1''$. 
We also exploit the following relation:
\be
a_4'=\frac{1}{2f-rf'} \left( rf''-\frac{rf'^2}{f}
+2f'-\frac{2f}{r} \right)a_4\,,
\ee
where $a_4$ is related to ${\cal H}$ by 
$a_4=(f/2)\sqrt{h/f}\,{\cal H}$. 
In the following, we derive the no-ghost conditions and radial propagation speeds for the dynamical perturbations $\psi$
and $\delta \phi$. We then apply these results to analyze the stability of BHs discussed in Sec.~\ref{BHsec}. Since the no-ghost conditions and radial speeds are sufficient to rule out all these BH solutions, we do not derive the angular propagation speeds in the even-parity sector. 
In GR without a scalar field $\phi$, we discuss the linear stability of planar BHs against perturbations in both radial and angular directions in Appendix~C. In this case, all the stability 
conditions are trivially satisfied.

\subsection{Region with $f>0$, $h>0$}

In the region where $f>0$ and $h>0$, 
the components of the matrix ${\bm K}$ determine 
the no-ghost conditions for dynamical perturbations 
$\psi$ and $\delta \phi$. They are given by 
\ba
\hspace{-0.8cm}
K_{11} &=& 
\frac{{\cal P} \mu^4 f^2 h^2}
{2\sqrt{fh}\,{\cal H}^2[2rf k^2 {\cal H}
+h \mu (rf'-2f)]^2}>0\,,\\
\hspace{-0.8cm}
{\rm det}\,{\bm K} &=& K_{11}K_{22}-K_{12}^2 
\nonumber \\
\hspace{-0.8cm}
&=& \frac{(2{\cal P}-{\cal F}){\cal F} 
\mu^4 fh}{4 {\cal H}^2 \phi'^2
[2rf k^2 {\cal H}+h \mu (rf'-2f)]^2}>0\,,
\ea
which translate to 
${\cal P}>0$ and $(2{\cal P}-{\cal F}){\cal F}>0$.
We recall that the absence of Laplacian instabilities 
in the odd-parity sector requires ${\cal F}>0$.
Under this condition, the ghosts in the even-parity 
sector are absent if
\be
2{\cal P}-{\cal F}>0\,.
\ee

To derive the radial propagation speeds, 
we assume the solution to the perturbation 
equations for $\psi$ and $\delta \phi$ 
in the form 
$\vec{\mathcal{X}}^{t}
=\vec{\mathcal{X}}^{t}_0 e^{-i (\omega t-k_r r)}$, 
where $\vec{\mathcal{X}}^{t}_0$ is a constant 
vector. In the limits of large $\omega$ and $k_r$, 
nonvanishing solutions to $\vec{\mathcal{X}}^{t}_0$
exist under the condition 
\be
{\rm det} \left|fh c_{r, {\rm even}}^2 {\bm K}
+{\bm G} \right|=0\,.
\label{deteq}
\ee
where $c_{r, {\rm even}}$ is the radial 
propagation speed measured in proper time 
$\tau=\int \sqrt{f}\,{\rm d}t$.
Solving Eq.~(\ref{deteq}) for 
$c_{r, {\rm even}}^2$, we obtain the 
following two solutions:
\begin{widetext}
\ba
\hspace{-0.5cm}
c_{r1,{\rm even}}^2
&=& \frac{{\cal G}}{{\cal F}}\,,\\
\hspace{-0.5cm}
c_{r2,{\rm even}}^2
&=& \frac{1}{(2{\cal P}-{\cal F})\mu^2 
\sqrt{fh}} 
\left[ \sqrt{\frac{h}{f}} (\mu-2r {\cal H}) 
( r^2 f'{\cal H}^2
+2rf {\cal G}{\cal H}-f\mu {\cal G} )
+ 4c_2 r^2 \phi' {\cal H}^2
+8c_3 h \phi' r^2 {\cal H} 
(\mu-r {\cal H}) 
\right]\,.
\label{cr2d}
\ea
\end{widetext}
To avoid Laplacian instabilities of even-parity 
perturbations in the radial direction, 
we require  $c_{r1,{\rm even}}^2>0$ and 
$c_{r2,{\rm even}}^2>0$.
We find that $c_{r1,{\rm even}}^2$ 
coincides with the value $c_{r,{\rm odd}}^2$ 
given by Eq.~(\ref{crodd1}). 
These correspond to the squared radial propagation 
speeds of the gravitational perturbations $\psi$ 
and $\chi$ in the even- and odd-parity sectors, 
respectively. Thus, under the linear stability 
conditions ${\cal G}>0$ and ${\cal F}>0$ in the 
odd-parity sector, we have $c_{r1,{\rm even}}^2>0$.
The squared radial propagation speed of 
the scalar-field perturbation $\delta \phi$ 
is given by $c_{r2,{\rm even}}^2$.

\subsection{Region with $f<0$, $h<0$}

In the region characterized by $f<0$ and 
$h<0$, the ghost-free conditions are 
determined by the matrix components of 
${\bm G}$, such that 
\begin{widetext}
\ba
G_{11} &=& \frac{rf h^2 \mu^2 
[\{ 4f\mu {\cal G}-2r^2 f'{\cal H}^2
- r {\cal H} (4f {\cal G} - f' \mu) 
\} \sqrt{fh}-4r f\phi'({\cal H}c_2
+ 2c_3 h \mu-2c_3 rh {\cal H})]}
{4{\cal H} [(2k^2 f{\cal H}+f'h \mu)r 
- 2f h \mu]^2}>0\,,\\
{\rm det}\,{\bm G} &=& G_{11} G_{22}-G_{12}^2 
\nonumber \\
&=& \frac{f^2 h^2 \mu^2 {\cal G}
\{ 4 {\cal H} [ 2 c_3 h (r {\cal H} 
-\mu)-{\cal H}c_2 ] r^2 \phi' \sqrt{fh} 
+ h (\mu-2r{\cal H})(r^2 f' {\cal H}^2
+ 2 rf {\cal G}{\cal H} -f \mu {\cal G})\} }
{4 {\cal H}^2 \phi'^2
[(2k^2 f{\cal H}+f'h \mu)r 
- 2f h \mu]^2}>0\,.
\ea
\end{widetext}
The radial propagation speeds for $\psi$ 
and $\delta \phi$ can be obtained by 
substituting $\vec{\mathcal{X}}^{t}
=\vec{\mathcal{X}}^{t}_0 e^{-i(\omega r-k_r t)}$ 
into their perturbation equations of motion.
In the large $\omega$ and $k_r$ limits, the 
dispersion relation is given by 
\be
{\rm det}\,\left| fh {\bm K}+
c_{r,{\rm even}}^2 {\bm G}
\right|=0\,,
\ee
where $c_{r,{\rm even}}$ denotes the radial 
propagation speed measured in proper time.
Solving this equation, we obtain two solutions 
for $c_{r,{\rm even}}^2$, given by 
\begin{widetext}
\ba
\hspace{-0.5cm}
c_{r1,{\rm even}}^2 
&=&
\frac{{\cal F}}{{\cal G}}\,,\\
\hspace{-0.5cm}
c_{r2,{\rm even}}^2 
&=& \frac{fh \mu^2 (2{\cal P}-{\cal F})}
{ h (\mu-2r{\cal H})(r^2 f' {\cal H}^2
+ 2 rf {\cal G}{\cal H} -f \mu {\cal G})
-4 {\cal H} [ {\cal H}c_2-2 c_3 h (r {\cal H} 
-\mu) ] r^2 \phi' \sqrt{fh}}\,.
\label{cre}
\ea
\end{widetext}
Both $c_{r1,{\rm even}}^2$ and 
$c_{r2,{\rm even}}^2$ must to be positive 
to avoid Laplacian instabilities in 
the radial direction.
Notice that $c_{r1,{\rm even}}^2$ corresponds to 
the squared radial propagation speed of 
the gravitational perturbation $\psi$. 
Indeed, this is equivalent to the value  
$c_{r,{\rm odd}}^2$ in the region $f<0$ and $h<0$. 
As long as the linear stability conditions 
${\cal F}>0$ and ${\cal G}>0$ in the odd-parity 
sector are satisfied, the Laplacian instability 
for $\psi$ is avoided. The scalar-field perturbation 
$\delta \phi$ has the squared radial propagation 
speed $c_{r2,{\rm even}}^2$ given by Eq.~(\ref{cre}).
Compared to the expression for $c_{r2,{\rm even}}^2$ 
in the region $f>0$ and $h>0$, where 
the term $2{\cal P}-{\cal F}$ appears 
in the denominator of Eq.~(\ref{cr2d}), 
the same term appears in the numerator
of Eq.~(\ref{cre}).

\subsection{Strong coupling problem}

We apply the stability conditions for 
even-parity perturbations to planar BHs 
discussed in Sec.~\ref{BHsec}. 
To this end, we use the background solution 
$\phi'(r)=1/r$ together with 
Eqs.~(\ref{heq}) and (\ref{feq}).

Let us consider Model~1 in the region where 
$f>0$ and $h>0$. Computing the quantity 
$2{\cal P}-{\cal F}$, we find
\be
2{\cal P}-{\cal F}=0\,,
\ee
at any distance $r$. This immediately 
implies that 
\be
{\rm det}\,{\bm K}=0\,.
\label{pro1}
\ee
Therefore, the BH solution with 
$\phi'(r)=1/r$ suffers from a strong 
coupling problem 
in any region where $f>0$ and $h>0$.
In Model~1, we also find the 
following property:
\be
{\rm det}\,{\bm G}=0\,,
\label{pro2}
\ee
along the solution $\phi'(r)=1/r$.
Since the terms inside the square brackets 
of Eq.~(\ref{cr2d}) are proportional to
${\rm det}\,{\bm G}$, the value of 
$c_{r,{\rm even}}^2$ is undetermined.

If we consider the region $f<0$ and $h<0$ 
in Model~1, we find that both properties 
(\ref{pro1}) and (\ref{pro2}) hold along 
the solution $\phi'(r)=1/r$. 
In this regime, the radial coordinate $r$ 
plays the role of time, so the condition 
${\rm det}\,{\bm G}=0$
implies that the strong coupling problem 
is also unavoidable. 
In summary, this BH solution suffers from an 
infinitely strong coupling problem at any value of $r$.

We have shown the relations (\ref{pro1}) and (\ref{pro2}) for Model~1. However, the same properties also hold for models with an arbitrary power $n$ along the solution 
$\phi'(r)=1/r$, namely, for the Horndeski coupling functions given in Eq.~(\ref{Gchoice}).
This implies that, irrespective of the choices 
of coefficients $c_n$, the strong coupling problem inevitably arises in all models constructed from the infinite sum of curvature corrections.
Moreover, this problem persists for all values of $r$, 
including $r=0$, where the background BH solution itself remains regular.
In this situation, nonlinear perturbations become uncontrollable, which makes the background BH solution physically illegitimate.
We stress that this strong coupling problem originates from the pathological behavior of the scalar-field perturbation $\delta \phi$ along the solution $\phi'(r)=1/r$.

%%%%%%%%%%%%%%%%%%%%%%%%%%%%%
\section{Conclusions} 
\label{consec}
%%%%%%%%%%%%%%%%%%%%%%%%%%%%%

In this paper, we have studied the linear stability of regular planar BHs in four-dimensional theories constructed by taking an infinite sum of Lovelock curvature invariants, along with a conformal 
rescaling of the metric.
The effective four-dimensional action belongs to a subclass of shift-symmetric Horndeski theories, characterized by the coupling functions 
$G_{2,3,4,5}(X)$ given in Eqs.~(\ref{G2})-(\ref{G5}). The convergence of these Horndeski functions depends on the choice of the coefficients $c_n$. 
We have considered two examples: $c_n=1$ (Model~1) 
and $c_n=[1-(-1)^n]/(2n)$ (Model~2), 
for which the corresponding coupling functions are given by Eqs.~(\ref{example1}) and (\ref{example2}), respectively.

In Sec.~\ref{BHsec}, we discussed how planar BH solutions 
regular at $r=0$ arise on the background given by 
Eq.~(\ref{line}). The common feature is that the scalar-field 
current $J$ must vanish everywhere to ensure the finiteness of $\phi'(r)$ at the horizon. 
In both Models 1 and 2, the solution to the scalar-field equation is given by $\phi'(r)=1/r$. This leads to metric components $f$ and $h$ that remain finite at $r=0$, 
see Eqs.~(\ref{fhso}) and (\ref{fhmoB}) for Models 1 and 2, respectively. 
Indeed, the leading-order term of the Kretschmann 
scalar expanded around $r=0$ has a finite value 
$24/\ell^4$ in both models. 

To study the linear stability of regular planar BHs, 
we consider perturbations on the background line element~(\ref{line}). As discussed in Sec.~\ref{persec}, the mode functions associated with the two-dimensional flat topology are given by cosine (or sine) functions of the form (\ref{eigen}). 
The perturbations can be decomposed into odd- and even-parity modes under rotations in the two-dimensional 
$(x_2,x_3)$ plane. 
We discussed the gauge transformation properties 
of each perturbed field and imposed the gauge condition given in Eq.~(\ref{gaugecon}) to fix the components of  the transformation vector $\xi^{\alpha}$.

In Sec.~\ref{oddsec}, we obtained the linear stability conditions for planar BHs under odd-parity perturbations in shift-symmetric Horndeski theories. 
As long as the three inequalities
${\cal G}>0$, ${\cal F}>0$, and ${\cal H}>0$ are 
satisfied, there are neither ghost nor Laplacian 
instabilities in both regions where $f>0$ and $f<0$. 
In both Models~1 and~2, we showed that the ghost-free condition for odd-parity perturbations is violated for regular BHs with $\phi'(r)=1/r$ near $r=0$.
Moreover, these same BH solutions are prone to Laplacian instability in the angular direction.

In Sec.~\ref{evensec}, we derived the no-ghost conditions and radial propagation speeds 
of even-parity perturbations for planar BHs
by taking the large $\omega$ and $k_r$ limits.
In the even-parity sector, there are two dynamical perturbations $\psi$ and $\delta \phi$ arising from 
the gravitational and scalar-field sectors, respectively.
For regular BHs with $\phi'(r) = 1/r$, we showed that the determinants of the $2 \times 2$ matrices 
${\bm K}$ and ${\bm G}$ in the second-order 
Lagrangian (\ref{evenact}) identically vanish for all values of $r$, independent of the signs of $f$ and $h$.
This indicates an infinitely strong coupling problem, thereby excluding the regular BHs as physically legitimate solutions. We note that this issue arises from the pathological behavior of the scalar-field perturbation $\delta \phi$, whose kinetic term vanishes along the solution $\phi'(r) = 1/r$.

We thus showed that the regular planar BHs in four spacetime dimensions, obtained by taking an infinite sum of Lovelock curvature invariants, suffer from both instabilities under odd-parity perturbations and a strong coupling problem in the even-parity sector. Since the strong coupling issue arises for the solution $\phi'(r)=1/r$ in models with each 
value of $n$, the same problem persists 
in models with the infinite sum over $n$, 
regardless of the choice of coefficients $c_n$.
Note that a similar strong coupling problem also 
arises for the scalar-field perturbation on the isotropic and homogeneous cosmological background \cite{Tsujikawa:2025eac}. It will be of interest to investigate whether the 
four-dimensional SSS BHs that may exist in the same theory can remain stable under odd- and even-parity perturbations. 
Furthermore, one can extend the linear stability analysis to the $2+1$-dimensional 
regular BTZ BHs arising from an infinite tower of curvature corrections~\cite{Fernandes:2025eoc}.
We leave these issues for future work.

%%%%%%%%%%%%%%%%%%%%%%%%%%%%
\section*{Acknowledgements}
%%%%%%%%%%%%%%%%%%%%%%%%%%%%

We are grateful to Hideki Maeda for valuable correspondence.
ST is supported by JSPS KAKENHI Grant Number 
22K03642 and Waseda University Special Research 
Projects (Nos.~2025C-488 and 2025R-028).

\section*{Appendix~A:~General Horndeski couplings}\label{sec:gen_act}
\renewcommand{\theequation}{A.\arabic{equation}} 
\setcounter{equation}{0}

In this Appendix~A, we discuss the generalization 
of the Lagrangians in Models~1 and 2.
The coupling $G_5(X)$ in Model~1 can be generalized 
to the following form 
\be
G_5(X)= -2 \ell^2 \left[ \frac{1-2\ell^4 X^2}
{(1-2\ell^2 X)^2}+\ln \left( \frac{2\ell^2 
|X|}{|1-2\ell^2 X|} 
\right) \right]\,,
\label{example1app}
\ee
where the other three coupling $G_2(X)$, $G_3(X)$, 
and $G_4(X)$ are the same as those given in Eq.~(\ref{example1}).
We can identify three distinct regions 
for the variable $X$: 
Region I, defined by $X<0$ (unbounded from below); 
Region II, where $0<X<1/(2\ell^2)$; and Region III,  
characterized by $X>1/(2\ell^2)$ (unbounded from above). 
Solutions may belong to combinations of these three regions. The choice of boundary/initial conditions determines 
the regions to which the solution belongs.

In Model~2, we can choose the following 
generalized coupling functions:
\ba
G_2(X) &=& -\Lambda+\frac{2X (28\ell^4 X^2-3)}
{(1-4\ell^4 X^2)^2} \nonumber \\
& &+\frac{3}{2\ell^2}
\ln\! \left| \frac{1+2\ell^2 X}{1-2 \ell^2 X}\right|\,,\\
G_5(X) &=& -\frac{2\ell^4 X}{1-4\ell^4 X^2}
-\frac{\ell^2}{2} \ln\! \left| \frac{1+2\ell^2 X}{1-2 \ell^2 X}\right|\,,
\label{example2app}
\ea
where the other two couplings $G_3(X)$ and $G_4(X)$  
are the same as those given in Eq.~(\ref{example2}).
For this second model, we can identify three distinct regions: Region I, defined by $X<-1/(2\ell^2)$; Region II, where  $-1/(2\ell^2)<X<1/(2\ell^2)$; and Region III, charactrized by $X>1/(2\ell^2)$. Once again, in this extension of the theory, the variable $X$ remains unbounded.

\section*{Appendix~B:~Coefficients of the even-parity 
second-order action}\label{AppenB}
\renewcommand{\theequation}{A.\arabic{equation}} 
\setcounter{equation}{0}

The coefficients of the second-order Lagrangian 
(\ref{Lageven}) in the even-parity sector 
are given by
\begin{widetext}
\ba
& &
a_1=\frac{fh}{2} \sqrt{\frac{h}{f}} \phi' 
\left[ G_{3,X} r^2 \phi'+4r (G_{4,X} 
-G_{4,XX}h \phi'^2) -h \phi' 
(3G_{5,X}-G_{5,XX}h \phi'^2)
\right]\,,\nonumber \\
& &
a_2=\sqrt{fh} \left( \frac{a_1}{\sqrt{fh}} 
\right)'-\left( \frac{\phi''}{\phi'}
-\frac{f'}{2f} \right)a_1
+\frac{r}{\phi'} \left( \frac{f'}{f}
-\frac{h'}{h} \right)a_4
=a_1'+\frac{J}{2}\,,\nonumber \\
& &
a_3=-\frac{1}{2}\phi' a_1-r a_4\,,\qquad
a_4=\frac{f}{2} \sqrt{\frac{h}{f}}{\cal H}\,,
\qquad 
a_5=-\frac{1}{2 \phi'} \sqrt{\frac{f}{h}} 
\left( {\cal H}'+\frac{{\cal H}}{r} 
-\frac{{\cal F}}{r} \right)\,,
\qquad 
a_6=-\frac{a_4}{2h}\,,\nonumber \\
& &
a_7=a_3'\,,\qquad 
a_8=a_4'+\left( \frac{1}{r}
-\frac{f'}{2f} \right)a_4\,,\qquad
b_1=\frac{a_4}{2f}\,,\qquad 
b_2=-\frac{2a_1}{f}\,,\qquad 
b_3=-\frac{J}{f}\,,\qquad
b_4=-\frac{2a_3}{f}\,,\nonumber \\
& &
b_5=-2b_1\,,\qquad 
c_1=-\frac{a_1}{fh}\,,\nonumber \\
& &
c_2=\frac{\phi'}{4} \sqrt{\frac{h}{f}}
[2f r^2(G_{2,X}-G_{2,XX} h \phi'^2) 
-r (rf' + 4 f) h \phi'(3G_{3,X}-G_{3,XX}h \phi'^2) 
- 4 (rf' + f) h ( G_{4,XXX} h^2 \phi'^4 \nonumber \\
& &\qquad 
- 6 G_{4,XX} h\phi'^2 + 3G_{4,X}) 
+ f' h^2 \phi' (G_{5,XXX} h^2\phi'^4 
- 10 G_{5,XX} h \phi'^2 + 15 G_{5,X})]\,,
\nonumber \\
& &
c_3=\frac{\phi'}{4r} \sqrt{\frac{h}{f}}
[2 G_{3,X}f r \phi' + 2 ( rf' + 2 f) 
(G_{4,X}-G_{4,XX} h \phi'^2)-f' h \phi' 
(3 G_{5,X}-G_{5,XX}h \phi'^2)]\,,\nonumber \\
& &
c_4=-h \phi' c_3-\frac{h}{2r} \sqrt{\frac{f}{h}}
{\cal G}-\frac{f'}{2f}a_4\,,\qquad 
c_5=\frac{\phi'f'}{8f}a_1+\frac{rf'}{2f}a_4
-\frac{1}{4}\phi' c_2+\frac{1}{2}rh \phi'c_3
+\frac{1}{4}h \sqrt{\frac{f}{h}}{\cal G}\,,\nonumber \\
& &
d_1=\frac{a_4}{2f}\,,\qquad d_2=2h c_3\,,
\nonumber \\
& &
d_3=\frac{\phi'}{2r^2} \sqrt{\frac{h}{f}}
[2 G_{2,X} fr^2- G_{3,X}r (rf' + 4f)h\phi' 
- 4 (rf' + f) h (G_{4,X}-G_{4,XX} h \phi'^2) 
+ f'h^2 \phi' (3 G_{5,X}-G_{5,XX} h \phi'^2 )],
\nonumber \\
& &
e_1=\frac{1}{\phi' fh} \left[ 
\left( \frac{f'}{f}+\frac{h'}{2h} 
\right) a_1-2a_1'+a_2-2rh a_5
\right]\,,\qquad 
e_2=-\frac{1}{2\phi'} \left( \frac{f'}{f}a_1
+2c_2+4r h c_3 \right)\,,
\ea
\end{widetext}
where the expression of $e_3$ is not shown.

\section*{Appendix~C:~Stability of the 
planar BH in GR}
\label{AppenC}
\renewcommand{\theequation}{A.\arabic{equation}} 
\setcounter{equation}{0}

In GR without a scalar field $\phi$, 
the Horndeski functions are given by 
\be
G_2=-\Lambda\,,\quad G_3=0\,,\quad 
G_4=\frac{1}{2}\,,\quad G_5=0\,.
\ee
Considering the asymptotically AdS spacetime with 
$\Lambda=-3/\ell_{\Lambda}^2<0$, the planar BH in GR
has the following metric components
\be
f=h=\frac{r^2}{\ell_{\Lambda}^2}
-\frac{2M}{r}\,,
\label{fhso0}
\ee
where $M$ is a constant. 
We discuss the linear stability of this planar 
BH against odd- and even-parity perturbations.

In the odd-parity sector, the dynamical perturbation 
corresponds to the field $\chi$, for which 
${\cal F}={\cal G}={\cal H}=1$. In this case, 
there is no ghost instability, and the propagation speeds 
of $\chi$ are luminal in both the radial and angular directions.

In the even-parity sector, the only dynamical perturbation 
is the field $\psi$, since the scalar field $\phi$ 
is absent. The reduced second-order Lagrangian is 
expressed in the form 
\be
{\cal L}_{\rm even}=K_{\psi} \dot{\psi}^2
+G_{\psi} \psi'^2+M_{\psi} \psi^2\,,
\ee
where 
\ba
K_{\psi} &=& \frac{r^2 h^2}{f} \sqrt{\frac{f}{h}}
\frac{1}{(k^2-3h-\Lambda r^2)^2}\,,\\
G_{\psi} &=& -r^2 h^3 \sqrt{\frac{f}{h}}
\frac{1}{(k^2-3h-\Lambda r^2)^2}\,,\\
M_{\psi} &=& -h \sqrt{\frac{f}{h}}
[h k^4 - \Lambda r^2 (\Lambda r^2 - h)k^2  
+ \Lambda r^2 (\Lambda^2 r^4 - 9 h^2)]
\nonumber \\
& &/(k^2-3h-\Lambda r^2)^3\,,
\ea
where we have not set $f=h$ for generality.
In the regime characterized by $f = h > 0$, 
the ghost-free condition $K_{\psi} > 0$ is 
automatically satisfied. This is also the case 
for the regime $f = h < 0$, in which the no-ghost 
condition $G_{\psi} > 0$ always holds. 

The squared propagation speed for $\psi$ along 
the radial direction is given by
$c_{r,{\rm even}}^2=-G_{\psi}/(fh K_{\psi})=1$ 
for $f=h>0$ and 
$c_{r,{\rm even}}^2=-fh K_{\psi}/G_{\psi}=1$ 
for $f=h<0$. 
Taking the large $k$ limit, the squared 
propagation speed for $\psi$ along the 
angular direction yields 
$c_{\Omega,{\rm even}}^2=-(r^2/f)
M_{\psi}/(k^2 K_{\psi})|_{k \to \infty}=1$ 
for $f=h>0$ and 
$c_{\Omega,{\rm even}}^2=r^2 h M_{\psi}/(k^2 G_{\psi})
|_{k \to \infty}=1$ 
for $f=h<0$. 
Thus, for high-momentum modes, the planar BHs 
in GR suffer neither ghost nor 
Laplacian instabilities.

\bibliographystyle{mybibstyle}
\bibliography{bib}

\end{document}